\PassOptionsToPackage{unicode}{hyperref}
\PassOptionsToPackage{hyphens}{url}
\documentclass[preprint,1p]{elsarticle}
\usepackage{geometry}
\usepackage{xcolor}
\usepackage{amsmath,amssymb}
\usepackage{iftex}
\usepackage[unicode,hidelinks]{hyperref}
\usepackage{bookmark}
\usepackage{caption}
\usepackage{subcaption}
\hypersetup{
  pdftitle  = {Graph Transformer-Based Flood Susceptibility Mapping},
  pdfauthor = {Vemula Sreenath; Filippo Gatti; Pierre Jehel}
}

\ifPDFTeX
  \usepackage[T1]{fontenc}
  \usepackage[utf8]{inputenc}
  \usepackage{textcomp} 
\else 
  \usepackage{unicode-math} 
  \defaultfontfeatures{Scale=MatchLowercase}
  \defaultfontfeatures[\rmfamily]{Ligatures=TeX,Scale=1}
\fi
\usepackage{lmodern}
\ifPDFTeX\else
\fi
\IfFileExists{upquote.sty}{\usepackage{upquote}}{}
\IfFileExists{microtype.sty}{
  \usepackage[]{microtype}
  \UseMicrotypeSet[protrusion]{basicmath} 
}{}
\makeatletter
\@ifundefined{KOMAClassName}{
  \IfFileExists{parskip.sty}{%
    \usepackage{parskip}
  }{
    \setlength{\parindent}{0pt}
    \setlength{\parskip}{6pt plus 2pt minus 1pt}}
}{
  \KOMAoptions{parskip=half}}
\makeatother
\usepackage{listings}

\usepackage{longtable,booktabs,array}
\usepackage{calc} 
\usepackage{etoolbox}
\makeatletter
\patchcmd\longtable{\par}{\if@noskipsec\mbox{}\fi\par}{}{}
\makeatother
\IfFileExists{footnotehyper.sty}{\usepackage{footnotehyper}}{\usepackage{footnote}}
\makesavenoteenv{longtable}
\usepackage{graphicx}
\makeatletter
\newsavebox\pandoc@box
\newcommand*\pandocbounded[1]{
  \sbox\pandoc@box{#1}%
  \Gscale@div\@tempa{\textheight}{\dimexpr\ht\pandoc@box+\dp\pandoc@box\relax}%
  \Gscale@div\@tempb{\linewidth}{\wd\pandoc@box}%
  \ifdim\@tempb\p@<\@tempa\p@\let\@tempa\@tempb\fi
  \ifdim\@tempa\p@<\p@\scalebox{\@tempa}{\usebox\pandoc@box}%
  \else\usebox{\pandoc@box}%
  \fi%
}
\def\fps@figure{htbp}
\makeatother
\usepackage{bookmark}
\IfFileExists{xurl.sty}{\usepackage{xurl}}{} 
\hypersetup{
  hidelinks,
  pdfcreator={LaTeX via pandoc}}

\date{}
\usepackage{pdfpages}

\begin{document}

\begin{frontmatter}

\title{Graph Transformer-Based Flood Susceptibility Mapping: Application to the French Riviera and Railway Infrastructure Under Climate Change}

\author[1]{Vemula Sreenath\corref{cor1}}
\ead{vemula.sreenath@centralesupelec.fr}

\author[1]{Filippo Gatti}
\author[1]{Pierre Jehel}

\affiliation[1]{%
  organization={Universit\'e Paris-Saclay, CentraleSup\'elec, ENS Paris-Saclay, CNRS, LMPS -- Laboratoire de M\'ecanique Paris-Saclay},%
  city={Gif-sur-Yvette},%
  postcode={91190},%
  country={France}%
}

\cortext[cor1]{Corresponding author}

\begin{abstract}
Increasing flood frequency and severity due to climate change threaten infrastructure and demand improved susceptibility mapping techniques. While traditional machine learning (ML) approaches are widely used, they struggle to capture spatial dependencies and have poor boundary delineation between classes. This study introduces the application of the graph transformer (GT) architecture for flood susceptibility mapping to the French Riviera using topography, hydrology, geography, and environmental data. GT incorporates watershed topology via Laplacian positional encoders and attention. The GT model achieved AUC‑ROC 0.9739, slightly below XGBoost (0.9853), but achieves superior spatial coherence with higher Moran's I 0.6119 and lower Geary's C 0.4729 versus XGBoost (0.4416; 0.5563) and random forest (0.4790; 0.5186). Feature importance revealed a consistency across models, with elevation, slope, distance to channel, and convergence index being the critical factors. Positional encodings captured some spatial patterns, but physical factors dominated feature importance. Future susceptibility was mapped for 2050 under multiple Representative Concentration Pathways (RCPs) using ensemble precipitation (5, 50, and 95) percentiles and projected LULC. Very‑high susceptibility area spans 3.22--26.83\% (RCP 2.6), 3.45--28.42\% (RCP 4.5), and 2.53--36.30\% (RCP 8.5), compared with 6.19\% under current conditions. Corresponding railway length within very‑high zones ranges 21.16--61.74\% (RCP 2.6), 22.24--63.50\% (RCP 4.5), and 21.36--67.35\% (RCP 8.5), compared to 35.61\% under current conditions. The resulting maps and quantified rail exposure provide actionable inputs for flood control, asset hardening, and multi‑hazard risk mitigation in the region.
\end{abstract}

\begin{keyword}
Flood susceptibility; Graph Transformer; Explainability; Climate and LULC projections; Railway infrastructure; French Riviera
\end{keyword}

\end{frontmatter}

\section{1. Introduction}
Floods are the third most damaging disaster after storms and earthquakes between 1995 and 2015, according to the United Nations Office for Disaster Risk Reduction estimates. Approximately 150,000 flood events were recorded, with 157,000 lives lost, affecting 2.3 billion people and causing damages of around US \$662 billion (Ghosh et al., 2022). Additionally, flooding causes annual damage of US\$14.6 billion to road and rail infrastructure globally, with 7.5\% of global railway assets exposed to 1 in 100-year flood events (Koks et al., 2019). In the context of the European Union, flood-related damage to railways amounts to €581 million annually (Bubeck et al., 2019), and 25\% of all the natural disaster events occurring to French railways correspond to flooding (Cheetham et al., 2016). Among railway infrastructure components, tracks are the most vulnerable to flood damage only after landslides, which are earthquake- or flood-triggered (Montoya-Araque et al., 2025). Of all natural hazards affecting railway infrastructure, 40\% of the impact is caused by track flooding, resulting in line closure, and 47\% is related to landslides, which could be earthquake- or flood-triggered (Montoya-Araque et al., 2025). Tracks are the most vulnerable railway elements affected by flooding (Toribio Diaz \& Vallhonrat Blanco, 2025). Furthermore, climate change influences flooding patterns, magnitude, and intensity, resulting in a 40\% increase in flood frequency (Nguyen, 2023). Climate change and increased urban settlements, especially along rivers, will exacerbate flood damage in the coming decades. Estimates project an annual flooding loss of US\$1 trillion by 2050 (Habibi et al., 2023).

Given the disastrous effects of present and future flooding on human lives and infrastructure, one approach to mitigate risk could be to predict floods: their spatial and temporal distribution, magnitude, duration, and other details (Hong et al., 2018). However, the flooding process is challenging to predict accurately as this process is dynamic and complex (Hong et al., 2018; Bui et al., 2019). Alternatively, researchers develop a flood susceptibility map as it is an essential step that assists local authorities in risk mitigation and managing future floods (Pham et al., 2020). Several flooding factors (e.g., elevation of the region, precipitation, soil type) cannot be controlled, but local authorities can prohibit constructions in highly susceptible flood areas, promote vegetation growth, and construct reservoirs to mitigate flood risk (Widya et al., 2024).

The flood susceptibility map indicates the likelihood of an area being affected by flooding using data from the watershed\textquotesingle s topographic, meteorological, geological, and environmental characteristics (Pham et al., 2020; Nguyen, 2023). Thus, unlike flood hazard analysis, which incorporates temporal characteristics of flooding, flood susceptibility focuses only on spatial distribution. Furthermore, almost all researchers assume these variables are static (e.g., Bui et al., 2019; Ghosh et al., 2022; Pradhan et al., 2023) and assessed for present climate and land use conditions.

The French Riviera has experienced several significant floods in the past decades: the 1994 and 2010 Var floods, the 2015 Côte d\textquotesingle Azur flood, and the 2020 Storm Alex flood. Storm Alex, which affected multiple countries (and this region was one of them), was particularly notable, resulting in damage of €2.7 billion (Ginesta et al., 2023). Prakash and Manconi (2021) identified over 1,249 landslides triggered by this flooding event using deep learning. Despite these devastating floods in this region, to the authors' knowledge, no previous flood susceptibility studies have been published.

Modeling the hydrological process can be broadly classified into three main approaches: conceptual, physical, and empirical (Bui et al., 2019). Empirical approaches are often called ``black-box'' models that use less data and can produce highly predictive models. Data-driven machine-learning (ML) approaches correspond to this class and have been significantly used, as they can learn nonlinear patterns from the data without fully understanding the underlying process (Islam et al., 2021). Despite not modeling the entire process, these approaches can produce more accurate models than traditional approaches with fewer computational resources and reduce the need for subjective expert weights (Nguyen, 2023).

Artificial neural network (ANN) (Bui et al., 2020), convolutional neural network (CNN) (Gao et al., 2024), decision trees (Khosravi et al., 2019; Chen et al., 2020), random forest (RF) (Chen et al., 2020; Abedi et al., 2022), support vector classification (SVC) (Choubin et al., 2018), transfer learning (Zhao et al., 2021), and eXtreme Gradient Boosting (XGBoost) (Abedi et al., 2022) are some of the ML approaches used for flood susceptibility. In recent years, researchers have either used ensemble hybrid approaches (Choubin et al., 2018; Islam et al., 2021) to overcome the limitations of one or several models and/or used them in conjunction with metaheuristic optimization techniques. Ant colony optimization, genetic algorithm, and swarm intelligence algorithms are commonly used metaheuristic techniques (Arabameri et al., 2022). Mosavi et al. (2018) and Bentivoglio et al. (2022) provide a detailed overview of different ensemble and metaheuristic algorithms used in flood susceptibility, which are thus not discussed here.

Bentivoglio et al. (2022) highlighted from their comprehensive survey that all the models were developed for regular grids (e.g., ANN and CNN), resulting in a lack of generalization and limited exploitation of patterns in the data. They suggested a geometric learning approach as a future direction to overcome the traditional ML model limitations. Graph neural networks (GNNs) belong to this class of geometric learning in which the nodes could represent flooding factors and edges could represent flow direction or distance between two points. There are many variants in GNN architecture, such as vanilla GNN, graph attention network, and graph transformer. Vanilla GNN considers message passing to learn the pattern from the data -- each node computes messages from all its neighbors, then all the messages are aggregated, and node representations are updated based on the aggregated messages (Gilmer et al., 2017). In the case of the Graph ATtention network (GAT) (Veličković et al., 2017), attention between different edges is also considered along with message passing. However, these two architectures do not consider the global positioning of nodes in the message passing. Graph transformer (GT) (Dwivedi \& Bresson, 2020) overcomes this limitation by incorporating positional encodings (PEs) into the node embeddings.

Wang et al. (2023) developed the GAT-based flood susceptibility model using fewer inputs and obtained better performance than traditional ML approaches. Zhang et al. (2024) noted that conventional deep learning models do not perform well for complex environments as they fail to capture correlations between landslide occurrence and susceptible environments accurately, and they developed a GT-based model for landslide susceptibility. Belvederesi et al. (2025) used graph architecture to represent landslide spatial information and a transformer for regression to capture landslide susceptibility accurately, and they concluded that their results were promising. Despite the GT model being used for other natural hazards, it has not yet been explored for flood susceptibility.

The limited explainability of ML and ensemble models is often criticized. Pradhan et al. (2023) studied the importance of features using the SHapley Additive exPlanations (SHAP) (Lundberg \& Lee, 2017) model. They observed that land cover and elevation are the two most important features affecting flood susceptibility. Wang et al. (2024) developed a novel residual logistic regression to achieve high accuracy and explainability at the same time. They observed that distance to roads and channels were the two most influential parameters affecting high and very-high susceptibility classes. Gao et al. (2024) used the SHAP model to explain their CNN model and observed that relative elevation is the most influential parameter. The above observations demonstrate that each watershed area behaves differently as different parameters dominate in each of them.

Limited research focused on the effects of climate change and land use and land cover (LULC) on flood susceptibility, as most research focuses on current conditions. Climate change can result in weather patterns that are reflected in increased annual precipitation, frequency of droughts and heat waves (Janizadeh et al., 2021). Rogers et al. (2025) studied the increase in future flood exposure due to climate and population changes. They attribute 21.1\%, 76.8\%, and the remaining 2.1\% to climate, population changes, and both. Hence, urban areas are sensitive to both climate and population changes. Global warming results in increased temperature, resulting in increased water-holding capacity in the atmosphere and thus, periods of increased precipitation and drought are observed (Trenberth, 2011). Representative Concentration Pathways (RCPs) are proposed based on future greenhouse gas concentrations for different emission scenarios. RCP 2.6 (very stringent climate action), 4.5 (intermediate scenario), 6 (higher intermediate scenario), and 8.5 (minimal climate action) are widely used projection scenarios, and they correspond to an increase in mean global temperature of 1.5 -- 2 °C, 2 -- 3 °C, 3 -- 4 °C, and 4 -- 5 °C, respectively by 2100 (Meinshausen et al., 2011). Janizadeh et al. (2021) studied the effects of climate change and land use for the year 2050 for Iran under RCP 2.6 and 8.5 scenarios. They observed that high and very high susceptibility classes increased by 11.55\% and 7.49\% for RCP 2.6 and 10.61\% and 5.67\% for RCP 8.5 scenarios, respectively. Gharakhanlou and Perez (2022) observed that the impact of climate change and LULC was limited in two watersheds in Canada, with a mere 2.5\% increase in areas of high and very-high susceptible classes for 2050 and 2080 years for RCP 4.5. Nguyen et al. (2024) observed a significant increase in future flood risk for high and very-high classes compared to current conditions under different RCP scenarios and forecasted LULC values for 2035 and 2050. These works highlight the need to assess the study area\textquotesingle s climate and LULC projections on flood susceptibility.

The following objectives are proposed based on the aforementioned literature review:

\begin{enumerate}
\def\labelenumi{\arabic{enumi}.}
\item
  Develop flood susceptibility models for the French Riviera using the GT model and compare its performance to conventional ML models and SHAP-based explainability to developed models.
\item
  Incorporate climate and LULC projections into the GT model and study its influence on flood susceptibility.
\item
  Understand regions of railway tracks that are at risk of flooding under current and future flood susceptibility as an application of the GT model.
\end{enumerate}

\section{2. Materials and Methods}

\subsection{2.1 Study Area Description}

The considered watershed corresponds to Alpes de Haute-Provence, Alpes-Maritimes, and Var departments in the French Riviera (southeast of France), with geographical extents from 6°45' to 7°53' east and 43°36' to 43°94' north as shown in Fig. 1. The watershed comprises catchments from the Var, Paillon, Siagne, and Argens rivers, with a total area of around 2650 km\textsuperscript{2}. The Alps mountains are to the north of the watershed, the Mediterranean Sea to the south, and Italy to the east. The mean annual precipitation in the watershed is 800 mm, distributed throughout the year; October and November are the peak seasons and July and August are the lowest precipitation. Flood inundation map is obtained from the publicly available catalogue interministériel de données (IDE) géographiques. The total inundation area is observed to be 172.6 km\textsuperscript{2} (6.5\% of the watershed area). Hydraulic modeling, satellite and aerial images, field observations, and other techniques were used to derive this inundation database. These values are used as the target variables in training the model. Fig. 2 shows the railway track in the considered watershed, which has a length of around 283 km. It can be observed that it runs along the coast, with most of its length located within the inundated area (around 97 km or 35\%). The national rail network track shapefile is obtained from the French railways\footnote{(https://ressources.data.sncf.com/explore/dataset/fichier-de-formes-des-voies-du-reseau-ferre-national/export/) (last accessed on 30 September 2024)}.

\subsection{2.2 Sampling Methodology}

An equal number of flooded and non-flooded points are to be considered to create an unbiased dataset for training (Zhao et al., 2019). Hence, to train the ML models such as ANN, SVC, RF, and XGBoost, 1247 points are considered from the flooded and 1247 from the non-flooded dataset. While this equal sampling approach is simplistic, future work could explore class imbalance techniques such as Synthetic Minority Oversampling Technique (SMOTE) and Adaptive Synthetic Sampling (ADASYN) to address sampling biases better. Of these points, 70\%, 15\%, and 15\% of the combined (flooded and non-flooded) datasets are used to train, validate, and test the ANN model, respectively. In contrast, 80\% for training and 20\% for testing from the combined dataset are used for SVC, RF, and XGBoost models with 5-fold cross-validation. 811 points (65\% from each set) from flooded and non-flooded sets were considered in the case of the GT model to reduce the computational cost. From these points, 70\%, 15\%, and 15\% points are considered for training, validation, and testing sets, respectively, and are shown in Fig. 2. This is a binary classification problem with values either flooded or non-flooded, with values assigned as 1 and 0, respectively.

\subsection{2.3 Flood Conditioning Factors}

Identifying the flood conditioning factors is a crucial step in a flood susceptibility study. After detailed literature review, the following were selected: topographic (elevation, slope, aspect, plan curvature, convergence index, TPI: topographic position index, TWI: topographic wetness index, TRI: topographic roughness index, STI: sediment transport index, slope length (SL) factor, SPI: stream power index), hydrological (drainage density, distance to channel/road/sea coast/rail, precipitation), environmental (LULC, several normalized difference indices), and geological (lithology, soil type) factors.

\subsubsection{2.3.1 Topographic Factors}

Topography describes the watershed\textquotesingle s physical shape, significantly impacts flooding patterns, and influences runoff, indirectly affecting precipitation (Bui et al., 2020; Islam et al., 2021). A 25m resolution DEM is obtained from the publicly available BD Alti, a government geosciences agency\footnote{https://geoservices.ign.fr/bdalti (last accessed on 30 September 2024)} for the considered departments. All the topographic factors are computed from the DEM model.

Elevation is an important factor directly affecting flooding and is inversely related. Lower areas are more prone to flooding than elevated areas, as water accumulates at lower elevations from the higher elevations (Islam et al., 2021). Consequently, coastal regions are more prone to flooding in the considered watershed. Fig. 3(a) shows the elevation levels of the considered area. It can be observed that the Alps Mountain ranges in the north and east have higher elevations with a peak value of 1845 m and relatively flat terrain along the coast.

The slope is another critical factor that directly affects flooding. Steeper slopes have increased runoff velocity and lower infiltration capacity, resulting in water accumulation downstream of the channel at gentle slopes (Nguyen, 2023). Fig. 3(b) shows slope values ranging from 0 to 1.31 radians (= 75°) with a mean value of around 12°. Gentle slopes are observed to be near the coast, and steeper slopes are in the inland mountainous regions. The convergence index resembles plan curvature, measuring water convergence (Habibi et al., 2023), as shown in Fig. 3(c). Details of aspect, plan curvature, flow accumulation, TWI, TRI, slope length, and SPI are provided in the supplementary.

\subsubsection{2.3.2 Hydrological Factors}

Hydrological factors describe water flow, distribution, and quantity in the terrain. Initially, channel networks are derived from the DEM, which is used for the subsequent analysis. Distance to the channel is a crucial factor affecting the amount of saturation and flooding. Regions closer to the channel are more prone to flooding, and farther distances are less prone to flooding (Islam et al., 2021). It is calculated as the Euclidean distance from the DEM to the nearest point on the derived channel network and is shown in Fig. 3(d) with a maximum distance of around 6,333 m. Details of distance to rail, coast, and road, as well as drainage density, are provided in the supplementary.

Precipitation is another major factor affecting flooding, as large amounts of precipitation observed during Storm Alex could cause significant flooding (Islam et al., 2021). Monthly precipitation data is obtained from the Météo-France website from January 1994 to September 2024\footnote{\url{https://meteo.data.gouv.fr/datasets/6569b3d7d193b4daf2b43edc}, (last accessed on 15 October 2024)}. Mean annual precipitation is computed from this data and then interpolated using the ordinary kriging method shown in Fig. 3(e). The minimum and maximum precipitation values of 652 mm and 943 mm are observed, respectively, with peak values in the inland mountainous regions and lower values along the coast. Precipitation interacts with elevation, slope, and proximity to the channel, which control water accumulation, runoff velocity and infiltration rates, and river exposure, respectively, and the runoff water could only cause flooding. So, precipitation in a region does not directly translate into flooding.

\subsubsection{2.3.3 Environmental Factors}

Environmental factors describe the impact of natural or human modifications affecting flooding with LULC and several normalized difference indices considered under this class. Fig. 3(f) shows a 10m resolution LULC raster map obtained from Karra et al. (2021). Permanent water bodies, trees (15 feet or taller) and dense vegetation, seasonally or all year-round flooded vegetation, crops and agricultural land, built areas (e.g., roads, railways, buildings), bare ground (areas with no vegetation), and rangeland are the classification categories present in the LULC. Each land cover category affects flooding differently. As discussed in the introduction, increasing urbanization and land use changes affect flooding significantly, especially in urban areas. Hence, it is an essential parameter that must be considered to understand its effect on current and future flooding. Areas in the vicinity of downstream water bodies are prone to flooding. Tall trees and dense vegetation significantly reduce water velocity and increase infiltration, resulting in decreased flooding. Most watershed areas correspond to trees and rangeland categories, especially in the inland Alps mountainous regions. Built areas create artificial barriers, preventing infiltration and increasing runoff and flooding. Built areas are mainly concentrated along coasts where cities such as Nice and Cannes are present. Low-lying, flat, bare-ground regions significantly contribute to flooding as they have no vegetation to break the water velocity. Bare grounds are present in the vicinity of the built areas. Details of different normalized difference indices are provided in the supplementary.

\subsection{2.4 Multi-Collinearity Analysis}

After computing the above topographic, hydrological, environmental, and geological inputs, it is essential to consider independent factors for training the ML model to get unbiased results (Habibi et al., 2023). Variance inflation factor (VIF) and tolerance (TOL) are widely used approaches to quantify linear correlations for all the inputs, with high VIF values indicating high multi-collinearity (e.g., Gao et al., 2024; Bui et al., 2019). VIF is computed using the inbuilt function from the \emph{statsmodels} library (Seabold \& Perktold, 2010), and TOL is computed as TOL = 1/VIF. VIF \textless{} 10 and TOL \textgreater{} 0.1 criteria are considered for the multi-collinearity. Table S1 shows the obtained VIF and TOL values with many collinear variables. When pairwise correlations (R) are considered, it is observed that elevation and precipitation are highly correlated (R = 0.835), agreeing with the previous observation that maximum precipitation occurred in the inland and minimum precipitation in the flat regions. Slope and TRI (R = 0.958), slope and TWI (R = -0.805), slope and STI (R = 0.873), TPI and profile curvature (R = 0.813), TRI and STI (R = 0.847), STI and SPI (R = 0.913), distance to rail and coast (R = 0.912), distance to coast and precipitation (R = 0.802), NDBI mean and NDVI mean (R = -0.815), NDVI mean and NDVI max (R = 0.941), NDVI mean and NWDI mean (R = -0.968), and NDVI max and NDWI mean (R = -0.894) are highly correlated. After removing highly correlated factors, the final flooding factors considered are shown in Table 1. It should be noted that despite precipitation and elevation strongly correlating with VIF \textgreater{} 10 and TOL \textless{} 0.1, precipitation is still considered as an input variable, as climate projections heavily influence it.

\section{3. Machine Learning Models}

\subsection{3.1 Model Details}

This section briefly overviews ML models: ANN, SVC, RF, and XGBoost. ANNs consist of hidden layers that are fully connected, and the model weights are trained using a backpropagation algorithm by iteratively minimizing the loss function. SVC (Cortes \& Vapnik, 1995) identifies an optimal hyperplane by maximizing the margin between classes in feature space. RF (Breiman, 2001) and XGBoost (Chen \& Guestrin, 2016) belong to the class of decision trees. RF constructs a parallel ensemble of decision trees using bootstrapping and aggregates the predictions using majority voting, thus reducing overfitting and improving generalization. XGBoost, on the other hand, is a sequential model that uses a gradient-boosting framework that iteratively builds trees to minimize residuals from previous trees and incorporates regularization techniques to prevent model overfitting.

The inputs and target are normalized using min-max scaling. \emph{Optuna} library (Akiba et al., 2019), an efficient framework designed for optimizing hyperparameters of ML models, was used for hyperparameter optimization. A subset of input features determined from the hyperparameter optimization is used to train the models because the model performance did not deteriorate when fewer inputs were used. The model predictions range from 0 to 1; if they are above 0.5, they are classified as flooded; otherwise, they are non-flooded. ANN model was trained using the \emph{Pytorch} library (Paszke et al., 2019), SVC and RF models were trained using the Scikit-learn (Pedregosa et al., 2011) library and XGBoost using the XGBoost (Chen \& Guestrin, 2016) library. Table S2 details the model architecture and input features considered for each model. It should be noted that no constraint was enforced to include precipitation and LULC as inputs, as these models are not used for future forecast predictions.

\subsection{3.2 Performance Metrics}

The following standard binary statistical metrics are used to assess the model performance as the flood susceptibility task is a classification problem: Area Under the Receiver Operating Characteristic Curve (AUC-ROC), sensitivity (true positive rate) (= TP/(TP+FN)), specificity (true negative rate) (= TN/(FP+TN)), Positive Predictive Value (PPV) (= TP/(TP+FP)), and Negative Predictive Value (NPV) (= TN/(TN+FN)). TP, TN, FP, and FN correspond to true positive, true negative, false positive, and false negative, respectively. Sensitivity and specificity are probabilities and measure the model's ability to correctly identify true positives (i.e., flooded regions) and true negatives (i.e., non-flooded regions), respectively. PPV and NPV indicate conditional probabilities: the probability that an area predicted as flood-susceptible is indeed flood-susceptible and non-flood-susceptible is indeed non-flood-susceptible, respectively (Habibi et al., 2023; Pradhan et al., 2023). ROC is a probability curve with 1 -- specificity (false positive rate) on the X-axis and sensitivity on the Y-axis. AUC represents the probability that the model will rank positive over negative samples (Wang et al., 2023). In general, AUC values range from 0.5 to 1, with a value of 1 implying that the classifier accurately predicts flooded regions as flooded and non-flooded areas as non-flooded. Conversely, an AUC value of 0.5 indicates performance like a random guess. Higher values of the considered metrics imply better model performance.

Standard binary metrics described above fail to capture how well a model delineates spatial boundaries between different classes, but provide overall model performance. Almost all the works in the literature identify model performance using binary classification metrics. In this work, Moran's I test (Moran, 1950) and Geary\textquotesingle s C (Geary, 1954) are additionally used to quantify the degree of spatial autocorrelation to understand whether similar susceptibility values cluster together spatially, or dispersed, or occur in a random pattern. Fenglin et al. (2023) research was one of the few works that considered Moran's I test on model residuals to validate spatial autocorrelation and choose the best-fit one. Gharakhanlou and Perez (2022) used Moran's I test to determine flooded and non-flooded points. However, these tests are used in the present work to understand whether spatial autocorrelations are captured in the model predictions.

A spatial weight matrix is constructed using a distance threshold (2000 m) criterion. Moran's I (Eq. 1) value ranges between -1 and 1. Moran's I value (Eq. 2) below E(I) implies a negative autocorrelation (dispersion) with a value of -1, implying perfect dispersion. Similarly, its value above E(I) indicates a positive autocorrelation (clustering), with a value of 1 implying an ideal clustering. Its value close to E(I) means randomness. Geary\textquotesingle s C (Eq. 5) ranges from 0 to values greater than 1, with an expected value of 1 under the null hypothesis of no spatial autocorrelation. Values significantly below 1 indicate positive spatial autocorrelation (clustering), while values significantly above 1 suggest negative spatial autocorrelation (dispersion). It can be observed that when neighborhood values deviate similarly from the mean, the product is positive in Eq. (1), increasing I. If they have similar values, the difference is minimum in Eq. (5), decreasing C. Additionally, these metrics are fundamentally dependent on the distance threshold ($\delta$). A smaller $\delta$ results in higher spatial autocorrelation, and a higher $\delta$ results in smaller values. Since floods typically influence a neighbourhood of 1-3 km, $\delta$ is chosen to be 2000m.

\begin{gather}
I \,=\, \frac{n\sum_{i}\sum_{j} w_{ij}\,\bigl(Y_{i}-\overline{Y}\bigr)\bigl(Y_{j}-\overline{Y}\bigr)}
               {\sum_{i}\bigl(Y_{i}-\overline{Y}\bigr)^{2}} \tag{1}\\[4pt]
E(I) \,=\, -\frac{1}{n-1} \tag{2}\\[4pt]
z_{i} \,=\, \frac{Y_{i}-\overline{Y}}{\sigma(Y)} \tag{3}\\[4pt]
y_{i} \,=\, \sum_{j} w_{ij}\, z_{j} \tag{4}\\[4pt]
C \,=\, \frac{n-1}{2}\,
        \frac{\sum_{i}\sum_{j} w_{ij}\,\bigl(Y_{i}-Y_{j}\bigr)^{2}}
             {\Bigl(\sum_{i}\sum_{j} w_{ij}\Bigr)\,\sum_{i}\bigl(Y_{i}-\overline{Y}\bigr)^{2}} \tag{5}
\end{gather}

In Eqs. 1 to 5, \emph{Y\textsubscript{i}} is the susceptibility value for location \emph{i}, \(\overline{Y}\) is the mean and $\sigma$ is the standard deviation, \emph{d\textsubscript{ij}} is Euleidean distance between i and j, \emph{w\textsubscript{ij}} is the spatial weights (= 1, if d\textsubscript{ij} \textless{} $\delta$; 0 if otherwise), n is number of spatial points, \emph{z\textsubscript{i}} is the standardized susceptibility value, \emph{y}\textsubscript{i} is the spatially lagged value, and E(I) represents the expected value of Moran's I under the null hypothesis of no spatial autocorrelation.

\subsection{3.3 Model Explainability}

Interpretability and explainability are commonly used to understand how the model predicts (Rudin, 2019). Interpretable models are transparent, allowing humans to understand their reasoning and predictions without additional tools (e.g., logistic regression and decision trees). On the other hand, explainable models require additional tools (e.g., SHAP) to understand the ``black box'' model, which humans cannot understand inherently (Pradhan et al., 2023). Rudin (2019) argued against using explainable models (or so-called statistics summaries) as they cannot represent the original black-box model perfectly, but the ideal scenario should be developing the interpretable ML models and acknowledging significant technical challenges in developing interpretable models. In explainable ML, there are two major approaches: global feature importance (e.g., using permutation) and local importance (e.g., SHAP). In the context of permutation (Fisher et al., 2019), an input feature is randomly permuted, and its effect on prediction is observed. The higher the change in model performance, the more critical the input feature is. Though it is simpler to implement, it does not inform anything about an individual prediction, whether an input feature positively or negatively affects the prediction or its dependency on features, but only for the generalized overall predictions.

SHAP provides a local explanation methodology using cooperative game theory (Shapley values) that attributes fair credit to each input feature based on its contribution to the model prediction. SHAP formulates the explanation as a linear model (Eq. 6) with Shapley values, considering all possible feature combinations and measuring each feature's marginal contribution. SHAP values satisfy four properties: efficiency (Shapley values sum to the difference between the actual and average predictions), symmetry (players with equal marginal contribution have the same score), the null player (player with zero marginal contribution results in zero reward), and additivity (two individual games can be added to obtain the overall contribution). This approach has significant advantages: local accuracy, missingness, consistency, and model-agnostic approximation (Lundberg \& Lee, 2017); however, its computation cost is expensive.

\begin{equation}
  g(z) = \phi_{0} + \sum_{i=1}^{M} \phi_{i} z_{i}' \, ,
  \label{eq:shap-linear}
\end{equation}

In Eq. 6, \(z' \in \left\{ 0,1 \right\}^{M}\), M is the total coalition size, \emph{$\phi$\textsubscript{i}} is the Shapley value.

\subsection{3.4 Results and Discussion}

This section discusses the results of the developed ML models. Table 2 shows the ML models' performance metrics for the test dataset. XGBoost achieved the highest performance across all metrics compared to other models with an AUC-ROC of 0.9853, demonstrating exceptional ability to discriminate between flooded and non-flooded classes. RF model exhibits strong performance that closely follows XGBoost. However, the highest sensitivity and specificity in the case of XGBoost imply that it predicts fewer false positives and negatives when predicting. SVC performs adequately but falls short compared to ensemble decision tree approaches such as XGBoost and RF, suggesting kernel-based approaches might struggle with high-dimensional, complex geospatial data. ANN significantly struggles to perform well compared to the remaining models. It achieved a modest AUC-ROC of 0.915, possibly due to limited training data size and necessitating a shallow feedforward network, which consequently struggles to learn flooded and non-flooded boundaries. Training the ANN with a metaheuristic optimizer might improve the model performance (Islam et al., 2021).

Discrete model predictions are rasterized using the Kriging approach, and the predictions are grouped into five classes using the Jenks natural break approach. It divides data into required groups by minimizing the variance within a group but maximizing the variance between groups. For uniformity in comparison, the GT model break ranges are used for all the models, though all the models have similar ranges. The break values are: 0 -- 0.0963, 0.0963 -- 0.258, 0.258 -- 0.456, 0.456 -- 0.7, and 0.7 -- 1 for very-low (cream color), low (tan color), moderate (green color), high (salmon color), and very-high (red color) classes, respectively. Additionally, values higher than 0.5 were considered flooded during the training. Consequently, high and very-high classes should ideally correspond to flooded regions, as the class break for moderate and high is around 0.456. Fig. S2, Fig. S3, Fig. 4, and Fig. 5 plot the flood susceptibility using the ANN, SVC, RF, and XGBoost model predictions and qualitative predictions are made using them. Table 3 shows the area of each susceptibility class for all the models.

It can be observed from Fig. S2 and Table 3 that the ANN model shows a good correlation between the channel networks and high and very-high susceptibility regions. It also captures topographic effects with high susceptibility regions in the lowland (near coasts) and low susceptibility in the mountainous areas. However, it fails to capture very-high susceptibility in the Var catchment and instead predicts high susceptibility. It predicts larger areas of moderate susceptibility in the Siagne catchment with blurred boundaries, failing to capture sharp transitions in flood risk. It also fails to predict very-high susceptibility in critical regions like the Var, Paillon, and Siagne catchments; instead, it only predicts high susceptibility. Furthermore, it likely overestimates the track section susceptible to flooding as it predicts that 72.85\% of track length falls within high and very-high susceptibility zones, significantly above the actual 35\%. This conclusion aligns well with Bentivoglio et al. (2022) discussion that convolution layers incorporate inductive bias and fully connected layers struggle to capture spatial correlations.

From Fig. S3 (supplementary), it can be observed that SVC shows better transitions and spatial accuracy between susceptibility zones than ANN predictions (e.g., Siagne Catchment). Regions classified as moderate are classified as low susceptibility by the SVC in the Siagne and Argens catchments. However, it slightly overpredicts the low susceptibility region and fails to identify very high susceptible regions in the Var catchment accurately. Generalized boundaries are still evident, especially when transitioning from low to very-low susceptibility regions. Around 51\% of the track length is estimated to be in high and very-high susceptibility regions.

RF model captures the transition from very-high to moderate to very-low susceptibility as we move away from channel networks in the Argens catchment, implying a better delineation of different regions than ANN and SVC, as shown in Fig. S4. This model identifies very-high susceptibility regions significantly better than previous models in all the catchments, though it fails to identify sharp boundaries precisely. High and very-high susceptible region areas add up to around 15.3\%. However, track length susceptible to flooding is slightly overestimated as the combined length of high and very-high regions exceeds 63\%.

XGBoost predictions are shown in Fig. 4, and despite its significantly higher AUC and other metrics, it struggles to predict very high susceptibility in the Var and Paillon catchments, but instead predicts high susceptibility because the predictions above 0.5 are treated as flooded regions and are not reflected in metrics. It significantly overpredicts low susceptibility regions (67\% area) and underpredicts very low susceptibility (1\% area) regions because the predictions below 0.5 are treated as non-flooded regions and are not reflected in metrics. Like SVC, it struggles to identify boundaries in the Argens and Siagne catchments precisely and predicts low susceptibility in the Argens catchment and moderate susceptibility in the Siagne catchment. XGBoost struggles to capture sharp boundaries precisely. Since it underestimates the very-high susceptibility region, it is reflected in the length of the track corresponding to this region. To conclude, despite XGBoost\textquotesingle s best performance in binary classification, it struggles significantly to differentiate between classes in flood susceptibility, and the metrics (e.g., AUC-ROC) poorly capture this.

Additionally, the SHAP plots of RF and XGBoost models are plotted (Fig. S5 supplementary) to determine the importance of features that influenced model predictions. It can be observed that topographic features such as slope, elevation, convergence index, and distance to channel (hydrological feature) are the most influential features in the RF model for predicting flooded and non-flooded classes. Whereas similar features in a different order are predicted in the case of the XGBoost model, indicating that the models have learned meaningful features to predict flood susceptibility. The SHAP values vary between -0.15 and 0.15 for the RF model (since magnitude is plotted, negative values are not evident) and -0.4 and 0.4. Additionally, SHAP values of influential features in the XGBoost are significantly high (e.g., elevation $\approx$ 0.4, slope $\approx$ 0.35), and the remaining features have minimal influence, which might be why XGBoost struggles with boundary delineation despite achieving the highest accuracy, i.e., overdependence on a few features.

Fig. S6 and Fig. S7 (supplementary) display the plots of Moran's I values for XGBoost and RF models, respectively. XGBoost model achieves Moran's I value = 0.4416 with Z-score = 230.2282 and Geary's C value = 0.5563 and Z-score = -252.8324 and RF model achieves Moran's I value = 0.4790 with Z-score = 252.1792 and Geary's C value = 0.5186 and Z-score = -274.2746, respectively. The p-value for all the tests was below 0.0001, and the z-score quantifying the statistical significance of observed spatial autocorrelation is significantly above 1.96 for a 95\% statistical level.

As discussed above, higher Moran's I and lower Geary's C values imply tighter clustering; consequently, the RF model exhibits slightly better spatial autocorrelations than the XGBoost model, resulting in better transitions from one susceptible class to another (which was concluded in the susceptibility plots as well). A higher scatter in the spatially lagged values in the case of the moderate and high susceptibility classes for the XGBoost model indicates that points from similar classes are less surrounded by points in these classes. Additionally, comparing high standardized values (above 1) and high spatially lagged values (above 1.5) of both these plots, it is evident that other high and very-high susceptibility areas consistently surround high and very-high susceptibility locations. In the next section, the GT architecture and its implementation are discussed.

\section{4. Graph Transformer Model}

To our knowledge, no transformer architecture (Vaswani et al., 2017) has been applied to the graph structure known as the graph transformer (Dwivedi \& Bresson, 2020) for the flood susceptibility problem; our idea with this manuscript is to create a blueprint that others could easily replicate to other regions with regional modifications.

\subsection{4.1 Graph Transformer Architecture}

The transformer architecture (Vaswani et al., 2017) developed at Google fundamentally revolutionized ML. Recurrent neural networks (RNNs) traditionally process one word at a time, significantly limiting scaling potential. Transformers introduced a self-attention mechanism where each word is connected to all other words, and attention learns how much to focus on each word. This significantly improved computation time because these operations could be parallelized and outperformed in capturing long-range spatial and temporal dependencies compared to RNNs. Since all the words are processed simultaneously, information on the position of word order is lost. Word positions are incorporated into input embeddings through PEs using sine and cosine functions to ensure each word has a unique representation. Then, they are projected to obtain queries (Q), keys (K), and values (V) with the hidden dimension size, and attention weights are computed using a scaled dot-product. Instead of learning single attention weights, they found it beneficial to learn multiple weight representations, as each attention could learn different representations of the same input, resulting in stable learning (Vaswani et al., 2017). Finally, the values from the attention layer are passed through a series of feedforward layers with residual connections after each layer to ensure smooth gradient flow. The transformer consists of encoder and decoder blocks; however, only the encoder block is used for our task, as discussed above.

Zhou et al. (2020) provided a detailed review of graph neural networks; interested readers could refer to that. A graph (G) consists of nodes (N) connected through edges (E) and is denoted as G(N, E). Unlike text data, graphs contain edge details and are irregularly connected, implying that nodes exist in non-Euclidean space (Zhou et al., 2020). Thus, despite the transformer\textquotesingle s success in natural language processing, it cannot be directly applied to the graph structure as it does not fully utilize the information in the graph structural topology and edges, and it is unclear what PEs mean when using sine and cosine. The GT architecture proposed by Dwivedi and Bresson (2020) comprises several aspects, such as the attention mechanism for a node being considered only between connected nodes and the Laplacian encoder instead of sine and cosine PE. They argued that Laplacian PEs (Eq. 7) generalize sine and cosine PEs for the graph structure.

\begin{equation}
\Delta \,=\, I - D^{-1/2} A D^{1/2} \,=\, U^{T}\Lambda U \tag{7}
\end{equation}

In Eq. 7, A is the adjacency matrix, I is the identity matrix, D is the degree matrix, U and $\Lambda$ are the eigenvectors and eigenvalues, respectively, from which the \emph{k} smallest non-trivial eigenvectors are obtained and used as PEs for a node (Dwivedi \& Bresson, 2020). The obtained PEs are concatenated with the input node features projected to the hidden dimension (H). Assuming multi-heads (\emph{n\textsubscript{heads}} number), the size of each head is thus d\textsubscript{k} = H/n\textsubscript{heads}. Scaled dot-product attention, A (Q, K), for a single head, \emph{i}, is computed using Eq. 8. The obtained attention scores are scaled by the edge weights (W\textsubscript{E}), given by Eq. 9 ($\odot$ corresponds to element-wise multiplication), which are used to compute the attention output (Eq. 10), and concatenated from all the heads to obtain final layer output. Finally, the obtained multi-head output is passed through feedforward layers with residual connections, followed by layer normalization after each layer (Zhang et al., 2024).

\begin{align}
A(Q,K) &= \operatorname{softmax}\!\left( \frac{QK^{T}}{\sqrt{d_{k}}} \right) \tag{8}\\
A'     &= (I + A) \odot W_{E} \tag{9}\\
Z_{i}  &= A' V \tag{10}
\end{align}

\subsection{4.2 Graph Construction}

In this section, details of the graph construction are discussed. The graph nodes are locations of flooded and non-flooded points considered, with the final flooding factors as node features. A transformer connects all possible nodes, forming a complete graph (Vaswani et al., 2017). However, it is computationally expensive and scales poorly as the size of the graph increases. Hence, instead of connecting each node with all other nodes, it only connects with similar node features (Zhang et al., 2024; Belvederesi et al., 2025). Thus, with local connections, GT can scale better for large watersheds. The principal components of node features are then obtained, retaining 95\% variance and cosine similarity between all the nodes, which is computed using the encoded principal component representation. Finally, each node is connected to itself, and the top k nodes (a hyperparameter) have the highest cosine similarity. It results in a directed graph. Edge weight could represent flow direction or proximity, but the cosine similarity value is used as edge weight in the present work. The flood susceptibility problem corresponds to a node-level prediction task -- the probability of a node being flooded. The binary cross-entropy loss function is considered, as it is a classification task.

\subsection{4.3 Graph Transformer Results}

The following are considered as hyperparameters: number of eigenvectors (num\_eigenvectors), number of neighbors each node is connected to (k-neighbours), number of multi-heads (num\_heads), hidden layer dimension (hidden\_dim), number of transformer layers, dropout to prevent overfitting, learning rate, and the dimension of the feedforward layer (ff\_multiplier) and optimized in Optuna. The obtained hyperparameters are shown in Table S2. Table 2 shows the GT model metrics. It can be observed that the GT model outperforms all models except XGBoost, with an AUC-ROC value of 0.9739, implying the model's superiority in discriminating flooded and non-flooded regions effectively and a specificity of 0.9426, implying that the model can accurately identify true positives. However, since the training data size is relatively small, decision tree methods perform slightly better than GT (Grinsztajn et al., 2022). Using datasets across multiple domains, Grinsztajn et al. (2022) demonstrated that decision tree models outperform deep learning approaches (e.g., GT) for training data of less than 10K samples. The GT model can leverage data effectively for large watersheds (e.g., national scale) and learn complex patterns better than traditional ML models. For instance, Lam et al. (2023) developed GraphCAST, a novel graph architecture-based global weather forecast, which outperforms deterministic simulations for 90\% of the target variables.

The Moran's I scatterplot for the GT model is shown in Fig. 5 with Moran's I value = 0.6119 (p-value \textless{} 0.0001) and Z-score = 161.5528, and Geary's C = 0.4729 (p-value \textless{} 0.0001) and Z-score = -300.2920. Since these values are better than those achieved by the RF and XGBoost models, the GT model could capture spatial autocorrelations better.

The area of each class and the length of the railway track in each class are shown in Table 3. As illustrated in Fig. 6, the GT model has improved spatial coherence in high-susceptibility zones compared to fragmented predictions from RF/XGBoost with more precise boundary delineation. Interestingly, the GT and RF models have similar susceptibility patterns, especially along the channel networks and the upper Alps mountainous regions. Additionally, the percentage of areas in these two zones of these two models was similar, implying a consistent identification of flood-prone areas. However, the length of the track in high and very-high susceptibility classes (51.2\%) is slightly lower than the RF model (63.29\%) but closer to the ground-truth flooded track length (35\%). The stark contrast between the GT and other models is its ability to sharply and coherently delineate the class boundaries compared to fragmented class boundaries in the RF model (especially evident for the low susceptibility in the Argens basin). The GT model slightly overpredicts moderate susceptibility in the mountainous regions and requires more computation (\textasciitilde{} 80 s for one Optuna run) and memory than the decision tree (\textless{} 10 s for one Optuna run) models. The GT model\textquotesingle s superior spatial delineation could be attributed to the graph-based attention mechanism and Laplacian position encoders.

The permutation approach (Fisher et al., 2019) was used on the GT model to understand the importance of global features, as shown in Fig. 7. Input features were permuted 100 times across bootstrap iterations to ensure robustness to obtain 2.5 and 97.5 percentiles, lower and upper bounds for 95\% confidence intervals. The importance values are normalized for better interpretation. The significance threshold is defined as the importance value if all the features are equal (=1/24) and is also plotted. It can be observed that elevation, distance to channel, slope, and convergence index are the top four critical features, with values of 0.22, 0.13, 0.12, and 0.11, respectively. The tight confidence bounds for all the features imply statistical confidence in the obtained results. These features are consistently ranked high in RF and XGBoost models, indicating that these might be the most critical features affecting flood susceptibility in this region, and all the models have learned meaningful features. Unlike XGBoost, the GT model does not strongly depend on a few features.

Additionally, lithology, NDBI, soil type, and the lowest eigenvector are close to the threshold. The considered Laplacian PEs indeed provide meaningful information; however, as we should expect, node features dominate and not the graph structure in importance. Interestingly, precipitation and LULC have a minor and negligible influence, respectively, on flood susceptibility. While current precipitation and LULC shows limited influence, their projected future values may amplify importance in scenario-based analyses.

We explore whether the Laplacian PEs can exploit the graph structure and provide meaningful information to the node features. In this regard, t-distributed stochastic neighbor embedding (t-SNE) (Van der Maaten \& Hinton, 2008) and Uniform Manifold Approximation and Projection (UMAP) (McInnes et al., 2020) techniques are used to understand PE embeddings, as shown in Fig. 8 with blue color for non-flooded and red for flooded. t-SNE is a nonlinear technique that preserves local structure - points close in the higher dimension will be close in the projected low-dimensional space (Van der Maaten \& Hinton, 2008). In contrast, UMAP balances to preserve local and global structures. The non-flooded and flooded nodes are separated, especially in the bottom left and right, respectively, of the t-SNE plot. Similarly, UMAP revealed partial clustering, especially the non-flooded nodes in the left part and flooded nodes in the bottom part of the plots. However, they are not separated in many regions, indicating that PEs could not fully distinguish between the flooded and non-flooded areas. A similar conclusion was drawn from Fig. 7, where flooding factors had higher importance than positional encodings.

Graph transformer parameters (k-neighbours, num\_eigenvectors, hidden\_dim, num\_heads, ff\_multiplier), dropout, and learning rate are considered for sensitivity analysis (Fig. S8) to evaluate their impact on both classification performance (AUC-ROC) and spatial autocorrelation quality (Moran\textquotesingle s I and Geary\textquotesingle s C). The One-Parameter-at-a-Time (OAT) approach is used to compute sensitivity as the maximum absolute deviation from baseline values. It can be observed that the transformer parameters (num\_heads and hidden\_dim) show the highest sensitivity, highlighting their importance in both classification performance and spatial autocorrelation quality. Graph construction parameters (num\_eigenvectors, k\_neighbours) have slightly lower sensitivity than transformer parameters. Training parameters (learning rate and dropout) show the lowest sensitivity. We also observe that Moran's I is more sensitive than Geary's C, perhaps because it is more sensitive in detecting subtle changes in spatial patterns. The following section uses the GT model to generate flood susceptibility maps for future scenarios.

\section{5. Flood Susceptibility for Future Scenarios}

Nguyen et al. (2024) observed a rapid increase in flood risk from 2005 to 2020 and predicted further increases by 2035 and 2050, attributed to changes in LULC and precipitation. This section examines the impact of future precipitation and LULC changes on flood susceptibility for the French Riviera. The Q05, Q50, and Q95 (5th, 50th, and 95th percentiles) from multi-model ensemble projected precipitation values were obtained for 2050 under RCP 2.6, RCP 4.5, and RCP 8.5 scenarios from the Drias portal\footnote{https://www.drias-climat.fr/ (last accessed on 15 October 2024)}. DRIAS represents a joint research collaboration between Météo-France, IPSL, CERFACS, and CNRM (Météo-France et al., 2025). Annual precipitation spans 375--890 mm (Q05), 650--1300 mm (Q50), and 890--1745 mm (Q95) across all RCPs, compared to current maximum mean annual values of 943 mm. Generally, increasing RCP from 2.6 to 8.5 raised ensemble precipitations, though some locations exhibited decreases. The prediction ranges demonstrate substantial variability across scenarios, with Q95 projections under RCP 8.5 reaching nearly double current precipitation levels. Additionally, projected LULC for 2050 was obtained from ArcGIS Living Atlas with 300 m resolution (Clark University, 2021), derived from ESA land cover data (2010--2018). Crop areas increased significantly from 1.23\% to 23.46\%, while other classifications slightly decreased.

All remaining flooding factors except projected precipitation and LULC were held constant---a simplification acknowledging that dynamic factors like topography and soil infiltration rates also evolve over time. Figs. 9 and S9 present flood susceptibility plots for Q50 under RCP 8.5 and Q95 under RCP 4.5 scenarios, respectively (for brevity). Model predictions based on median precipitation (Q50) under RCP 8.5 exhibit similar susceptibility patterns to current conditions; however, very-low susceptibility areas decreased while all other classes expanded. Substantial changes occur under Q95 RCP 4.5 projections, with significant areas classified as high and very-high susceptibility, consistent with previous studies (Janizadeh et al., 2021; Nguyen et al., 2024). Monte Carlo dropout was applied during inference, enabling uncertainty estimation through 100 stochastic forward passes. Epistemic uncertainty values (Fig. 10) remained below 0.034, with higher values near high and very-high susceptible zones.

Table 4 presents area percentages and railway track lengths for each susceptibility class under different 2050 RCP-quantile scenarios. Since Q05 projections across all RCPs fall significantly below current precipitation (652--943 mm), estimated areas and track lengths are correspondingly reduced. Q50 projections, with maximum values exceeding current estimates, show modest increases in high and very-high class areas and track lengths. However, Q95 projections reveal substantial increases in very-high susceptibility areas: 26.83\%, 28.42\%, and 36.3\% under RCP 2.6, RCP 4.5, and RCP 8.5, respectively, compared to 6.19\% under current conditions. Corresponding track lengths reached 61.74\%, 63.5\%, and 67.35\% for these scenarios versus 35.61\% currently. The RCP 4.5 scenario yielded slightly lower areas and lengths than RCP 2.6, reflecting underlying precipitation trends.

Finally, we assess the factors affecting this significant increase in flood susceptibility, as shown in Fig. S10 (supplementary), which is plotted for the future estimated data. It can be observed that the importance of elevation, slope, convergence index, and distance to the channel decreased using the forecasted data. Since annual precipitation values increased, its importance significantly increased. Similarly, increased precipitation directly affects the drainage density and lithology, resulting in their importance. The importance of NDBI could have also increased flood susceptibility in urban areas corresponding to the Siagne catchment. Interestingly, LULC has a limited effect on the model\textquotesingle s performance; perhaps the 300 m coarse resolution was insufficient. Finally, the importance of PEs significantly increases, implying that the spatial relationships are more critical for future scenarios.

\section{6. Conclusions}

The present research addresses a critical gap in flood risk management for the French Riviera -- a region experiencing multiple disastrous floods in the past decades. Our work introduces the graph transformer-based flood susceptibility analysis that outperforms traditional ML approaches and provides a quantitative railway vulnerability for the current and future scenarios.

We initially performed flood susceptibility analysis using ML approaches frequently used in the literature with topographic, hydrological, geological, and environmental factors. XGBoost model had the highest values in all the metrics, followed by random forest. However, when susceptibility maps were qualitatively verified, the RF model identified classification boundaries more effectively than the XGBoost model. SHAP plots provided insight into why this could happen: XGBoost's overreliance on a few parameters was not the case with the RF model. However, both models suggested elevation, slope, distance to channel, and convergence index as the influencing parameters, indicating that the models could learn actual flood processes. Spatial autocorrelations further validated the above conclusion: XGBoost had Moran's I and Geary's C values = 0.4416 and 0.5563, whereas RF model achieves 0.4790 and 0.5186, respectively, indicating that the RF model learned better spatial autocorrelations, resulting in better transitions from one susceptible class to another. Since flood classification is a multi-classification problem, we concluded that binary metrics cannot fully explain the model\textquotesingle s performance. Despite the RF model\textquotesingle s superiority over XGBoost, it fails to identify sharp boundaries precisely.

We proposed that GT learns spatial features better than traditional ML approaches, as they are incorporated into the node features. We used Laplacian PEs and demonstrated that they could identify flooded and non-flooded decision boundaries using t-SNE and UMAP techniques. However, PEs' influence on model prediction was limited as flooding factors dominated the model prediction. Interestingly, the flooding factors influencing RF and XGBoost models were the top four factors for the GT model. GT Susceptibility map had sharp and abrupt class boundaries with Moran's I and Geary's values 0.6119 and 0.4729, respectively, significantly better than the XGBoost and RF models, indicating that the GT model could effectively delineate the class boundaries. Additionally, track length corresponding to very-high class was 35.61\%. Our GT framework serves as a replicable blueprint that others could replicate in other regions with potential regional adaptation needs.

Finally, the effect of climate (using Q05, Q50, and Q95 precipitation percentiles) and LULC change was assessed for 2050 with the GT model under RCP 2.6, 4.5, and 8.5. For the very-high class, ranges (area and track length) are: RCP 2.6 = 3.22--26.83\% and 21.16--61.74\%; RCP 4.5 = 3.45--28.42\% and 22.24--63.50\%; RCP 8.5 = 2.53--36.30\% and 17.80--67.35\%; other classes generally increased relative to current conditions. The importance of precipitation, NDBI, and drainage density increased using forecasted data. Interestingly, LULC had limited importance (possibly due to 300 m coarser resolution), and PEs had increased importance, indicating spatial relations are important for future projections.

The major limitation of the present work on railway tracks is that the obtained flooding probability does not directly translate into track flooding since track elevation is ignored. Training a GT model is computationally challenging compared to traditional ML models. However, for a given region, the graph architecture could be precomputed, and inference time is significantly reduced, potentially enabling its integration with hydraulic models and use for real-time flood management. The outcome of our work can inform authorities in infrastructure planning and mitigation strategies.

\section*{Data availability}
The data that support the findings of this study are available from publicly accessible sources as cited in the manuscript; all datasets used are publicly available. 

\section*{CRediT authorship contribution statement}
Vemula Sreenath: Conceptualization; Methodology; Software; Validation; Formal analysis; Investigation; Data curation; Writing – original draft; Visualization. 
Filippo Gatti: Conceptualization; Methodology; Resources; Writing – review \& editing; Project administration. 
Pierre Jehel: Conceptualization; Resources; Writing – review \& editing; Project administration; Funding acquisition.

\section*{Declaration of generative AI and AI-assisted technologies in the writing process}
During the preparation of this work the authors used Claude Sonet 4.0 to improve readability and correct language. After using this tool, the authors reviewed and edited the content as needed and take full responsibility for the content of the publication.

\section*{Acknowledgments}
The authors acknowledge and appreciate the contributions of several organizations for their efforts and for making data publicly available: Météo-France for current precipitation values, SNCF for railway tracks, géographiques IDE for inundation maps, and IGN for DEM. The DRIAS climate services portal provided the climate projections data used in this study, which Météo-France and the national climate scientific community (IPSL, CERFACS, CNRM) implemented. Clark University developed the forecasted LULC data used in this study in partnership with Esri.

\section*{Funding}
This research was carried out in the Minerve project, supported by the French government within the France 2030 framework. The authors gratefully acknowledge this support.

\section*{Declaration of competing interest}
The authors declare that they have no known competing financial interests or personal relationships that could have appeared to influence the work reported in this paper.

\begin{table}[htbp]
  \centering
  \caption{Variance inflation factor (VIF) and Tolerance (TOL) for the final flooding inputs}
  \label{tab:vif_tol}
  \begin{tabular}{@{}llll@{}}
    \toprule
    S. No. & Feature                      & VIF    & TOL   \\
    \midrule
    0  & Elevation                    & 4.099  & 0.244 \\
    1  & Slope                        & 6.180  & 0.162 \\
    2  & Aspect                       & 4.634  & 0.216 \\
    3  & Plan Curvature               & 1.092  & 0.916 \\
    4  & Convergence Index            & 1.722  & 0.581 \\
    5  & Topographic Position Index   & 1.634  & 0.612 \\
    6  & Slope Length Factor          & 3.877  & 0.258 \\
    7  & Drainage Density             & 7.687  & 0.130 \\
    8  & Stream Power Index           & 2.341  & 0.427 \\
    9  & Distance to Road             & 1.990  & 0.502 \\
    10 & Distance to Channel          & 3.610  & 0.277 \\
    11 & Mean Annual Precipitation    & 32.422 & 0.031 \\
    12 & LULC                         & 4.313  & 0.232 \\
    13 & Soil Type                    & 10.131 & 0.099 \\
    14 & NDBI max                     & 1.459  & 0.685 \\
    15 & Lithology                    & 2.408  & 0.415 \\
    \bottomrule
  \end{tabular}
\end{table}

\begin{table}[htbp]
  \centering
  \caption{Performance metrics of the ML and GT models for the testing dataset}
  \label{tab:perf_test}
  \begin{tabular}{@{}llllll@{}}
    \toprule
    Model    & AUC-ROC & Sensitivity & Specificity & PPV    & NPV    \\
    \midrule
    ANN      & 0.915   & 0.8333      & 0.8138      & 0.8453 & 0.8205 \\
    SVC      & 0.9527  & 0.9000      & 0.8840      & 0.8858 & 0.8984 \\
    RF       & 0.9788  & 0.9320      & 0.9280      & 0.9283 & 0.9317 \\
    XGBoost  & 0.9853  & 0.9440      & 0.9400      & 0.9402 & 0.9438 \\
    GT       & 0.9739  & 0.9426      & 0.9264      & 0.9274 & 0.9417 \\
    \bottomrule
  \end{tabular}
\end{table}

\begin{table}[htbp]
  \centering
  \caption{The area of each susceptibility class and the track length in each class}
  \label{tab:class_area_length}
  \begin{tabular}{@{}llrrrrr@{}}
    \toprule
    Model & Metric & Very low (\%) & Low (\%) & Moderate (\%) & High (\%) & Very high (\%) \\
    \midrule
    ANN     & Area    & 29.30 & 29.44 & 23.07 & 13.83 &  4.36 \\
            & Length  &  1.00 &  8.95 & 17.20 & 39.88 & 32.97 \\
    SVC     & Area    & 28.15 & 39.55 & 19.52 &  9.94 &  2.85 \\
            & Length  &  3.64 & 26.35 & 18.67 & 26.73 & 24.61 \\
    RF      & Area    & 37.85 & 29.44 & 17.40 &  9.20 &  6.11 \\
            & Length  &  2.14 & 18.53 & 16.04 & 23.44 & 39.85 \\
    XGBoost & Area    &  0.78 & 67.15 & 22.31 &  7.73 &  2.03 \\
            & Length  &  0.41 & 14.92 & 30.62 & 37.91 & 16.14 \\
    GT      & Area    & 52.95 & 19.30 & 12.48 &  9.07 &  6.19 \\
            & Length  & 22.87 & 14.37 & 11.56 & 15.59 & 35.61 \\
    \bottomrule
  \end{tabular}
\end{table}

\begin{table}[htbp]
  \centering
  \caption{Area of each susceptibility class and the track length in each class for 2050}
  \label{tab:area_length_2050}
  \begin{tabular}{@{}lccccc@{}}
    \toprule
    Variable              & Very-low (\%) & Low (\%) & Moderate (\%) & High (\%) & Very-high (\%) \\
    \midrule
    RCP2.6, Q05 (Area)    & 76.58 & 10.60 & 5.60 & 4.00 & 3.22 \\
    RCP4.5, Q05 (Area)    & 73.89 & 12.14 & 6.21 & 4.31 & 3.45 \\
    RCP8.5, Q05 (Area)    & 74.82 & 12.47 & 6.22 & 3.96 & 2.53 \\
    RCP2.6, Q50 (Area)    & 45.36 & 18.53 & 13.09 & 11.19 & 11.83 \\
    RCP4.5, Q50 (Area)    & 53.06 & 16.37 & 11.26 & 9.40 & 9.91 \\
    RCP8.5, Q50 (Area)    & 38.94 & 19.24 & 14.40 & 13.22 & 14.20 \\
    RCP2.6, Q95 (Area)    & 20.08 & 17.36 & 17.04 & 18.69 & 26.83 \\
    RCP4.5, Q95 (Area)    & 23.85 & 15.77 & 15.23 & 16.73 & 28.42 \\
    RCP8.5, Q95 (Area)    & 16.35 & 13.59 & 15.00 & 18.76 & 36.30 \\
    RCP2.6, Q05 (Length)  & 42.46 & 11.69 & 10.45 & 14.24 & 21.16 \\
    RCP4.5, Q05 (Length)  & 38.74 & 12.06 & 11.41 & 15.55 & 22.24 \\
    RCP8.5, Q05 (Length)  & 41.34 & 10.63 & 13.91 & 16.32 & 17.80 \\
    RCP2.6, Q50 (Length)  & 21.04 & 10.72 & 7.23  & 14.09 & 46.92 \\
    RCP4.5, Q50 (Length)  & 25.89 & 8.82  & 9.22  & 13.69 & 42.37 \\
    RCP8.5, Q50 (Length)  & 20.06 & 11.10 & 6.77  & 12.25 & 49.83 \\
    RCP2.6, Q95 (Length)  & 8.52  & 8.72  & 10.35 & 10.67 & 61.74 \\
    RCP4.5, Q95 (Length)  & 7.50  & 5.59  & 12.15 & 11.25 & 63.50 \\
    RCP8.5, Q95 (Length)  & 5.98  & 4.85  & 9.66  & 12.17 & 67.35 \\
    \bottomrule
  \end{tabular}
\end{table}

\begin{figure}[htbp]
  \centering
  \includegraphics[width=\linewidth]{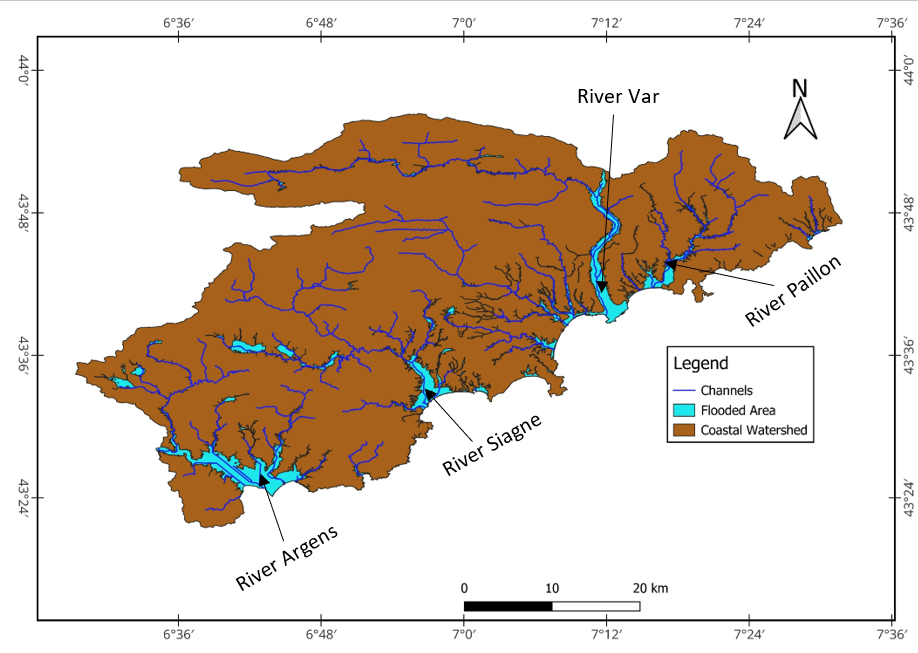}
  \caption{Inundation map of the French Riviera, along with channels derived from DEM}
  \label{fig:inundation}
\end{figure}

\begin{figure}[htbp]
  \centering
  \includegraphics[width=\linewidth]{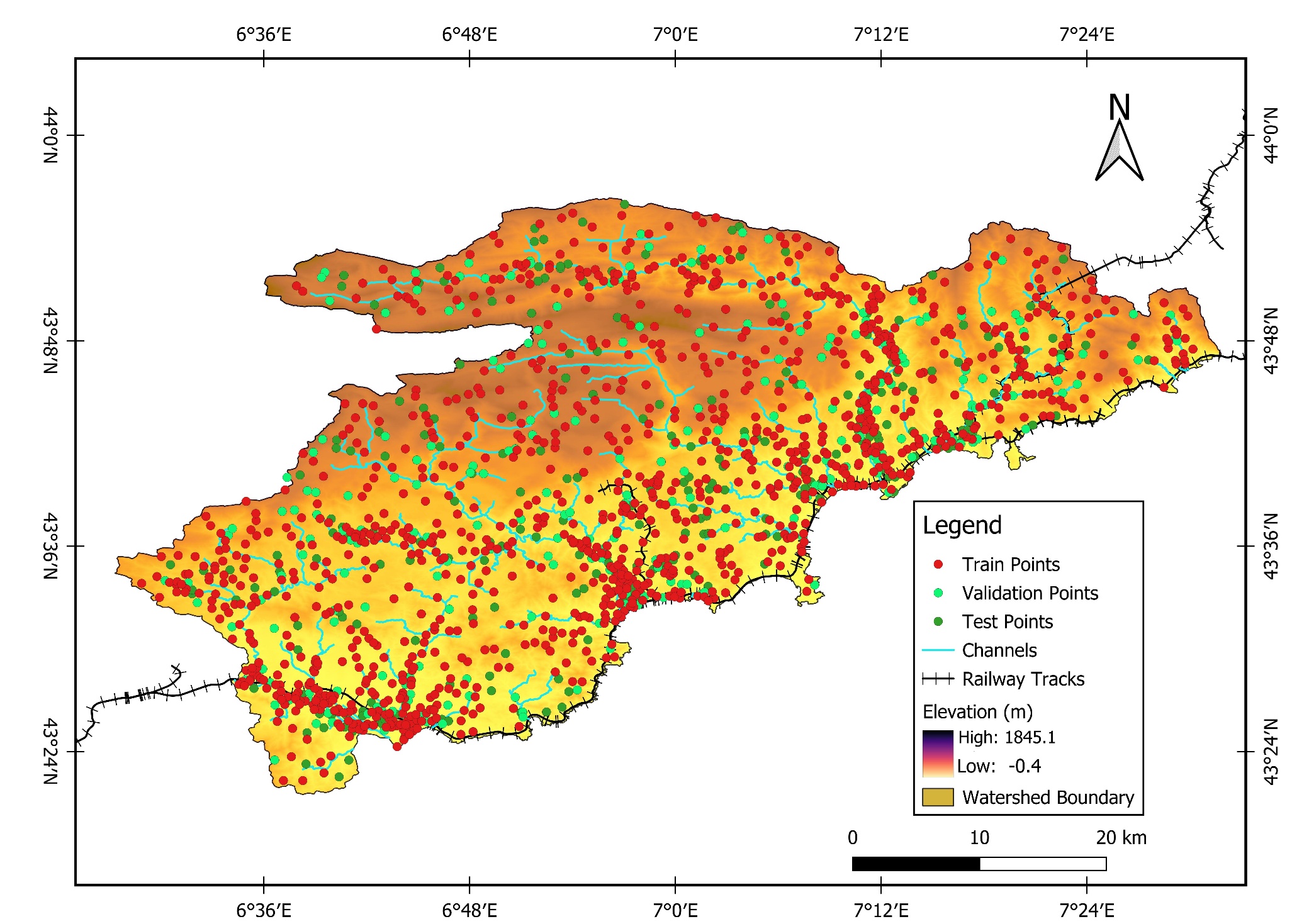}
  \caption{Training, validation, and testing points for the GT model and railway track}
  \label{fig:gt-splits-rail}
\end{figure}

\begin{figure}[htbp]
  \centering

  \begin{subfigure}[t]{0.48\textwidth}
    \centering
    \includegraphics[width=\linewidth]{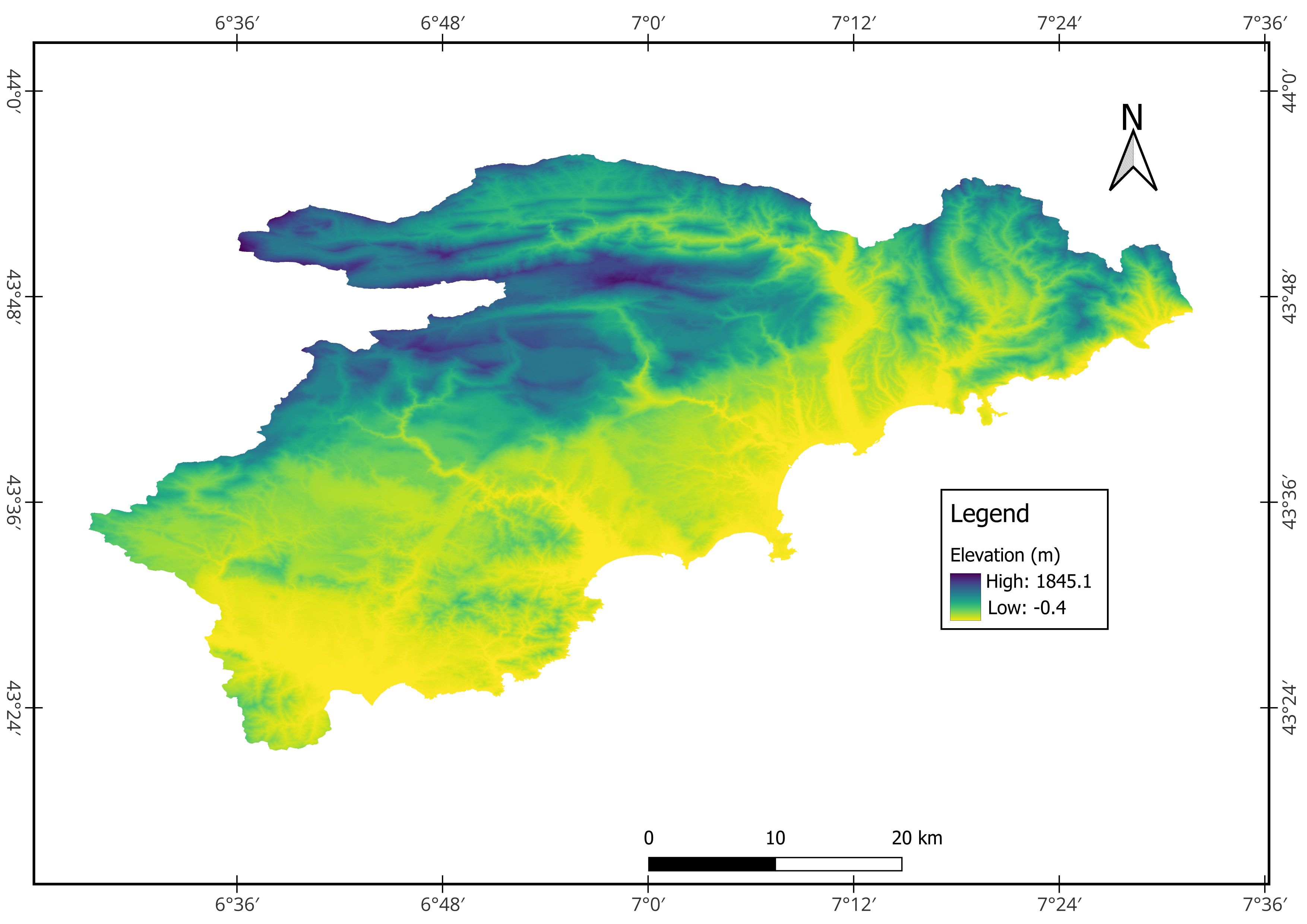}
    \caption{Elevation}
    \label{fig:3a}
  \end{subfigure}\hfill
  \begin{subfigure}[t]{0.48\textwidth}
    \centering
    \includegraphics[width=\linewidth]{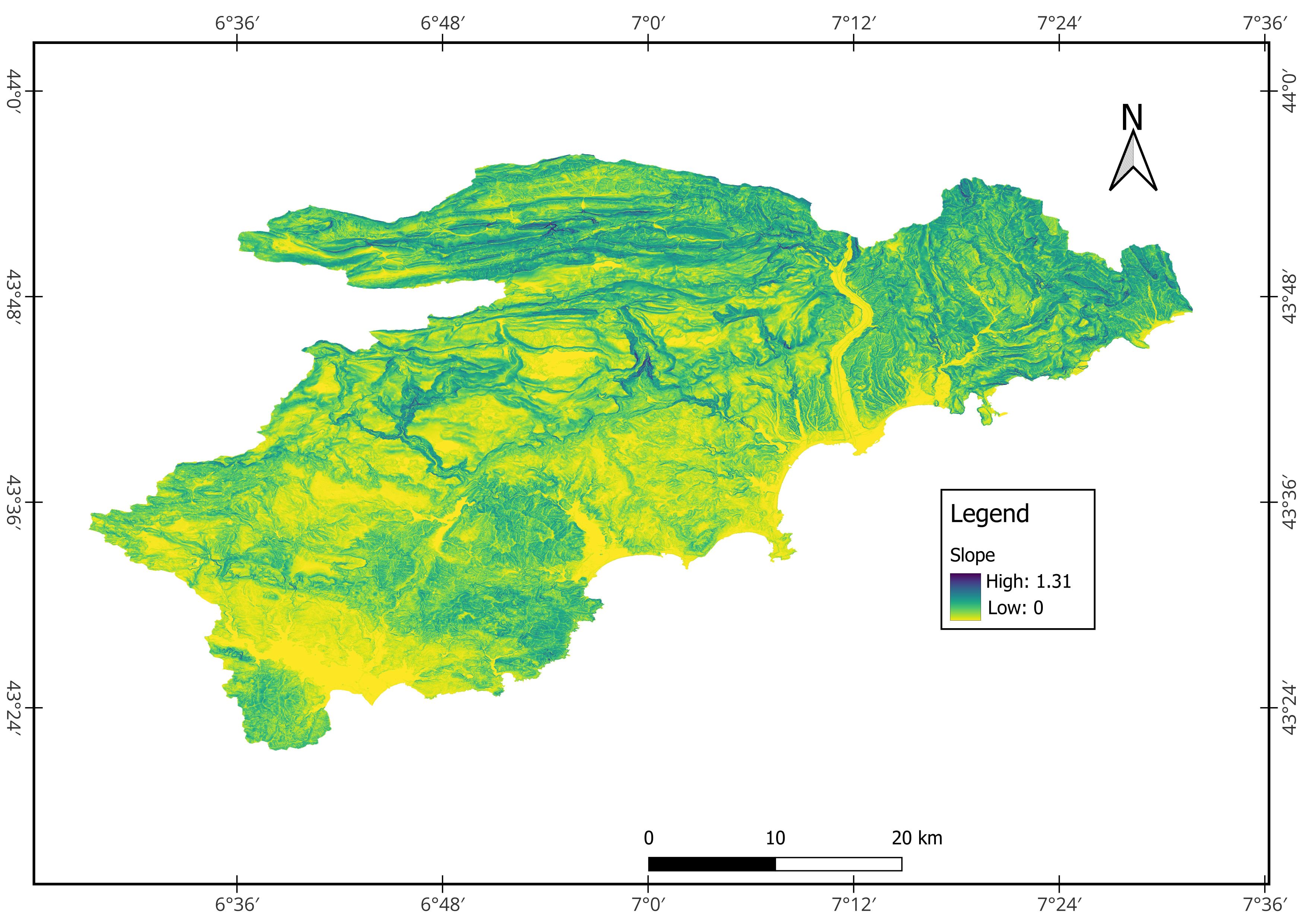}
    \caption{Slope}
    \label{fig:3b}
  \end{subfigure}

  \vspace{0.6em}

  \begin{subfigure}[t]{0.48\textwidth}
    \centering
    \includegraphics[width=\linewidth]{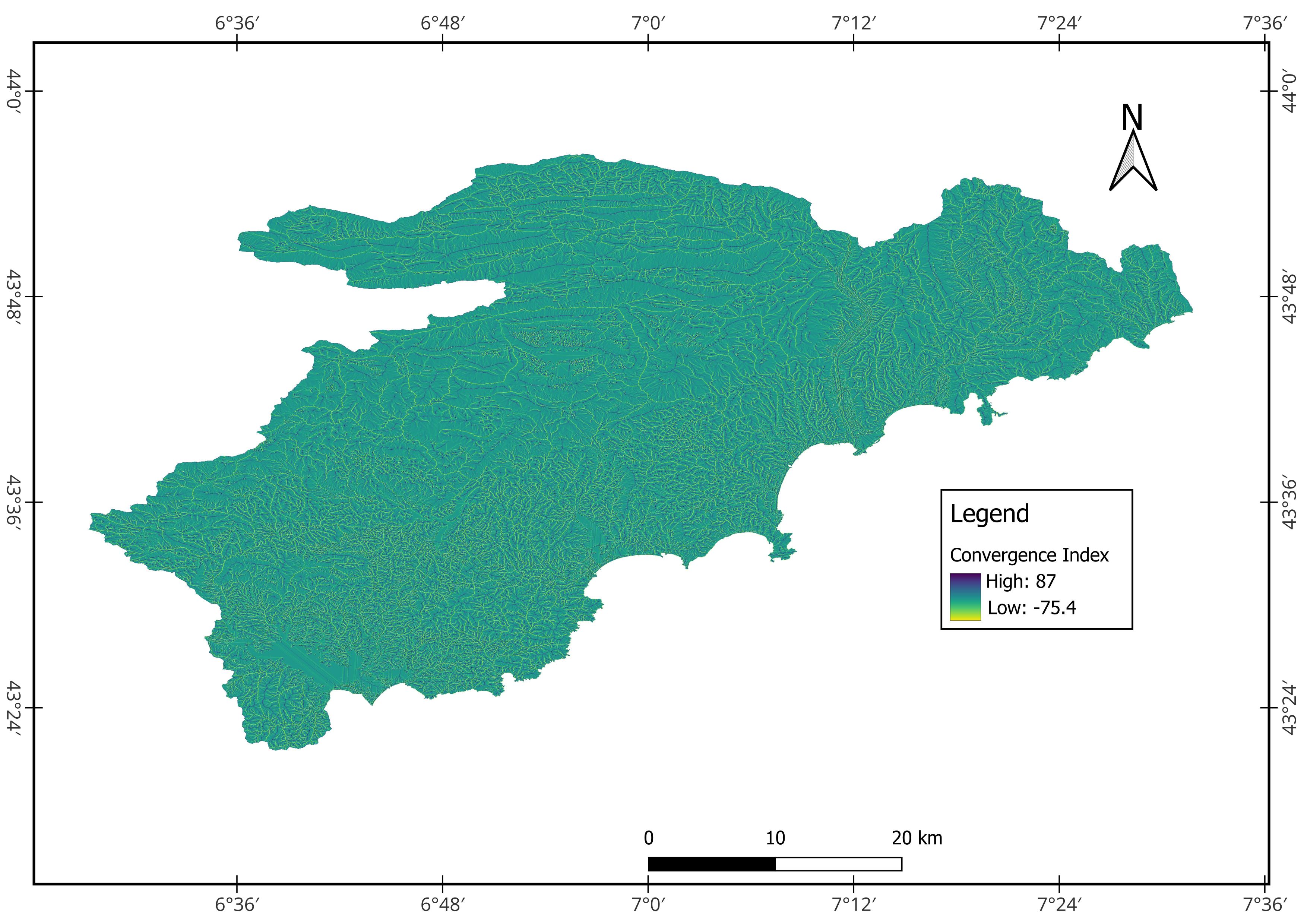}
    \caption{Convergence index}
    \label{fig:3c}
  \end{subfigure}\hfill
  \begin{subfigure}[t]{0.48\textwidth}
    \centering
    \includegraphics[width=\linewidth]{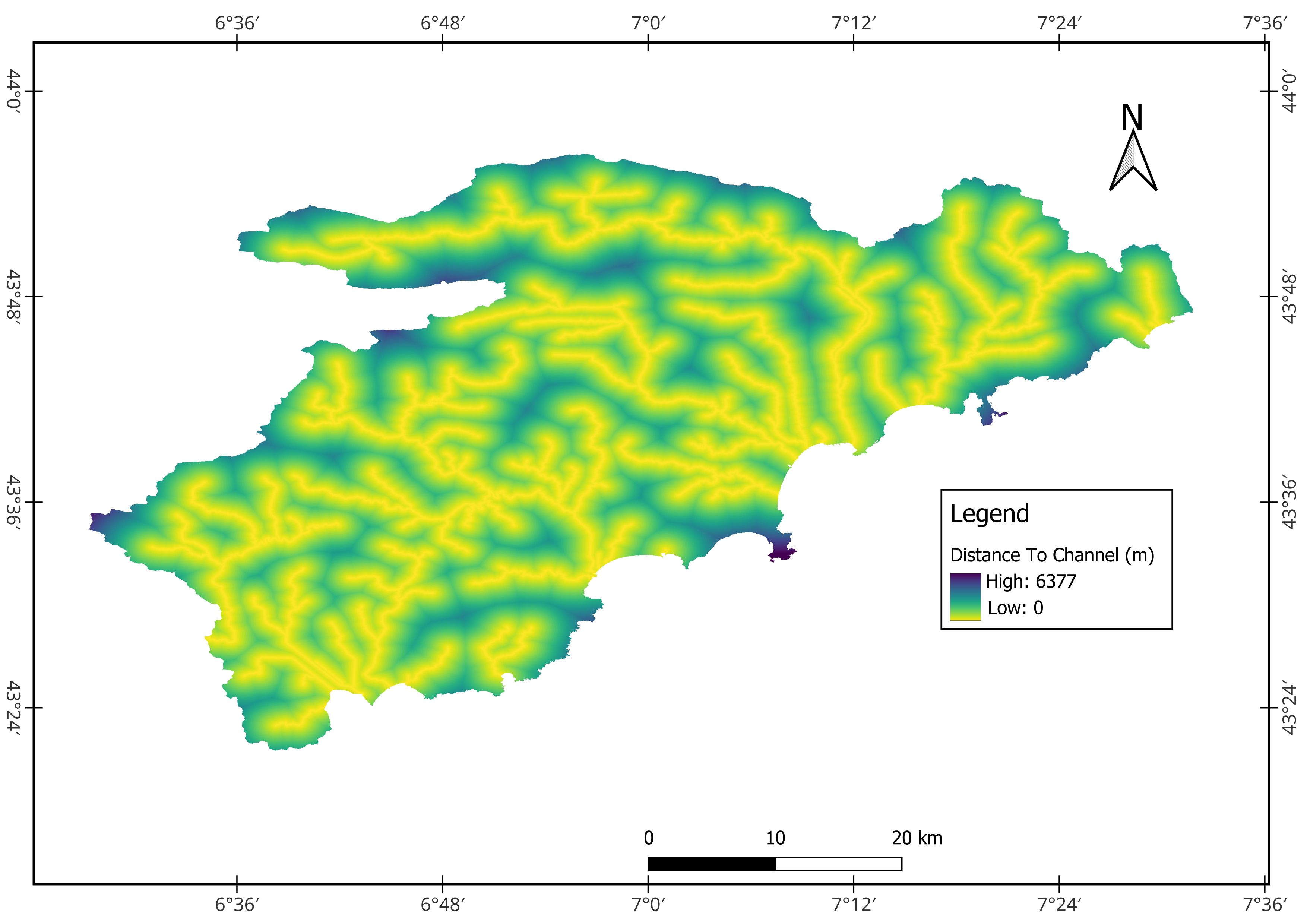}
    \caption{Distance to channel}
    \label{fig:3d}
  \end{subfigure}

  \vspace{0.6em}

  \begin{subfigure}[t]{0.48\textwidth}
    \centering
    \includegraphics[width=\linewidth]{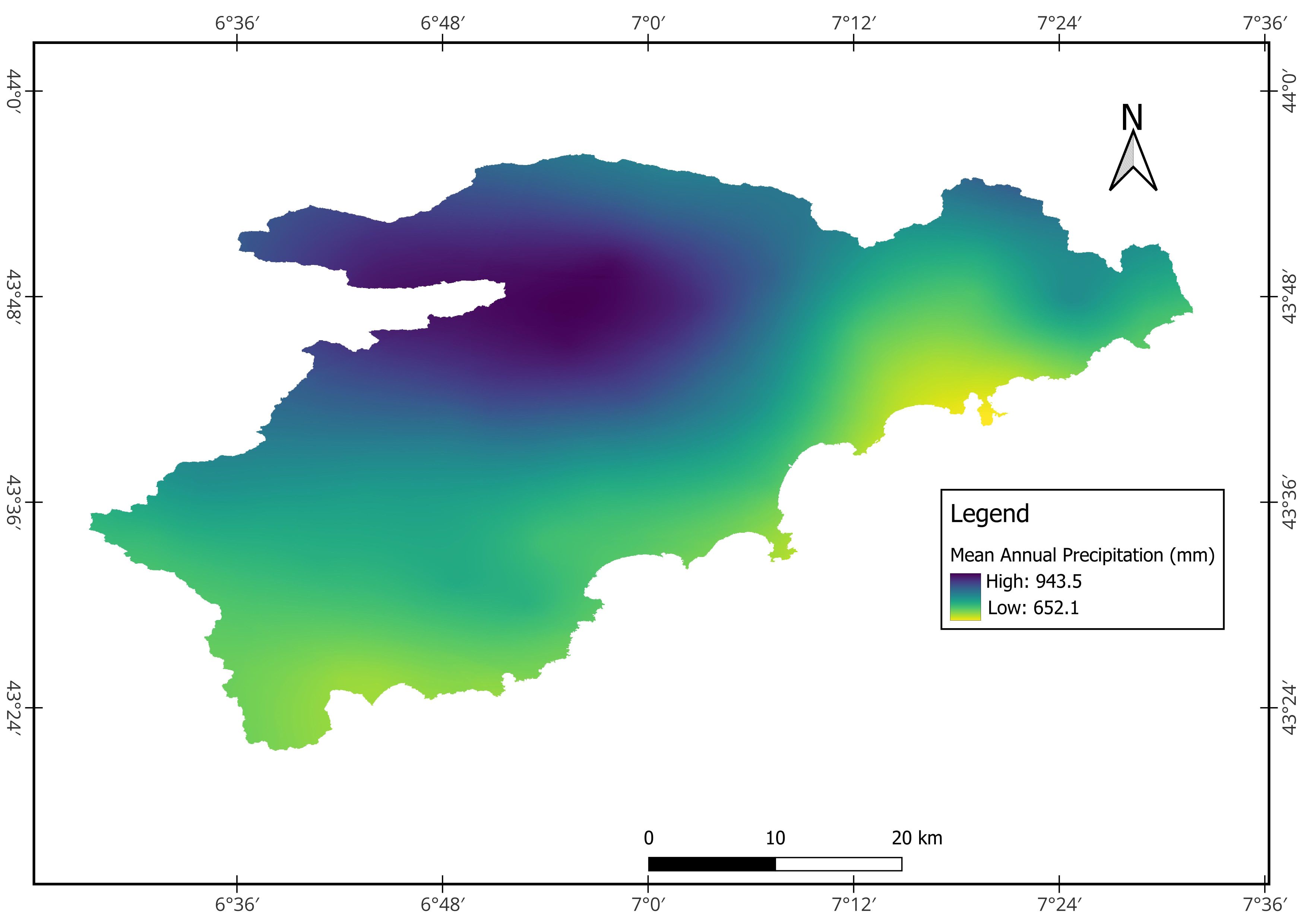}
    \caption{Mean annual precipitation}
    \label{fig:3e}
  \end{subfigure}\hfill
  \begin{subfigure}[t]{0.48\textwidth}
    \centering
    \includegraphics[width=\linewidth]{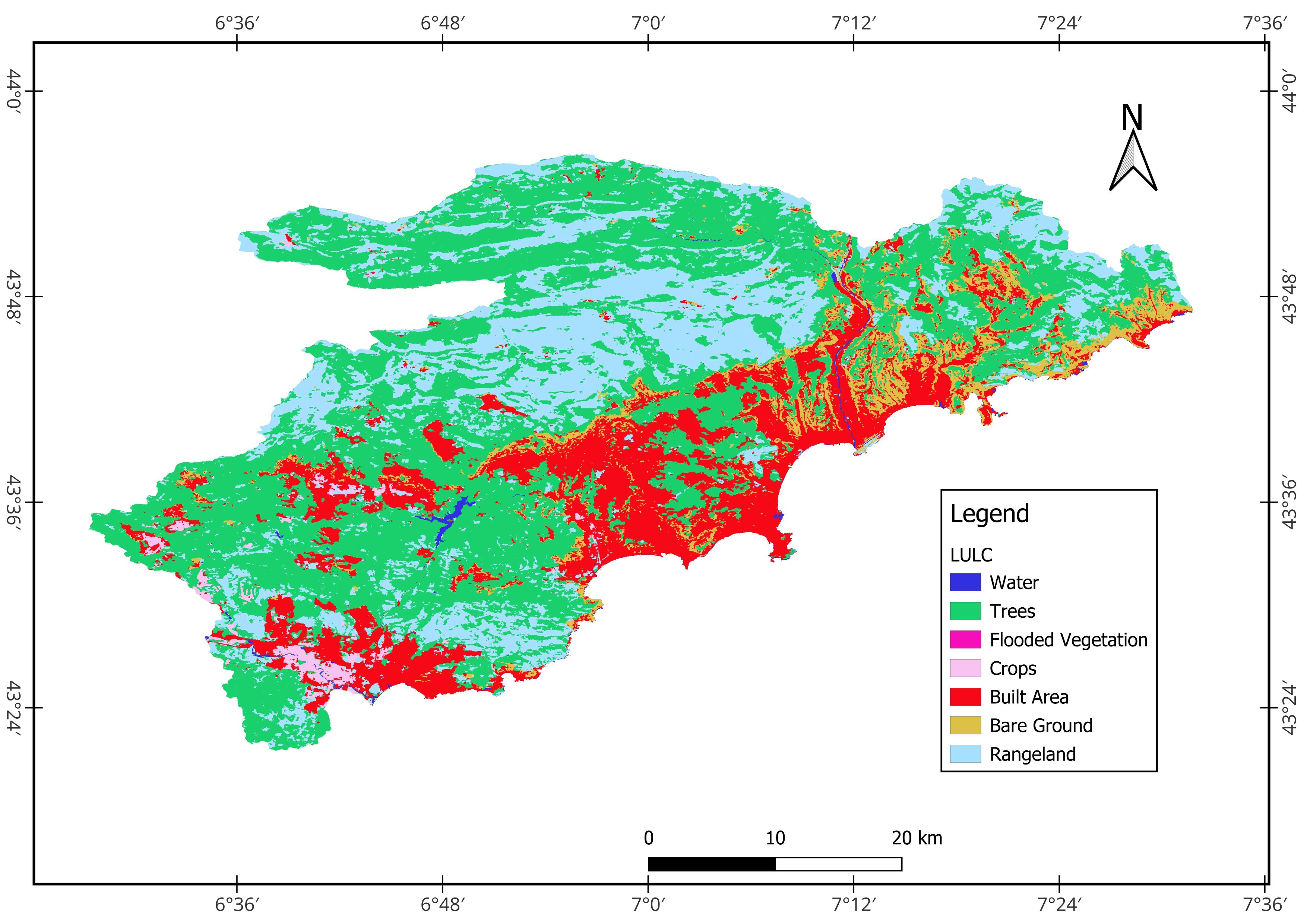}
    \caption{LULC}
    \label{fig:3f}
  \end{subfigure}

  \caption{Flood influencing factors: (a) elevation, (b) slope, (c) convergence index, (d) distance to channel, (e) mean annual precipitation, (f) LULC}
  \label{fig:inputs}
\end{figure}

\begin{figure}[htbp]
  \centering
  \includegraphics[width=\linewidth]{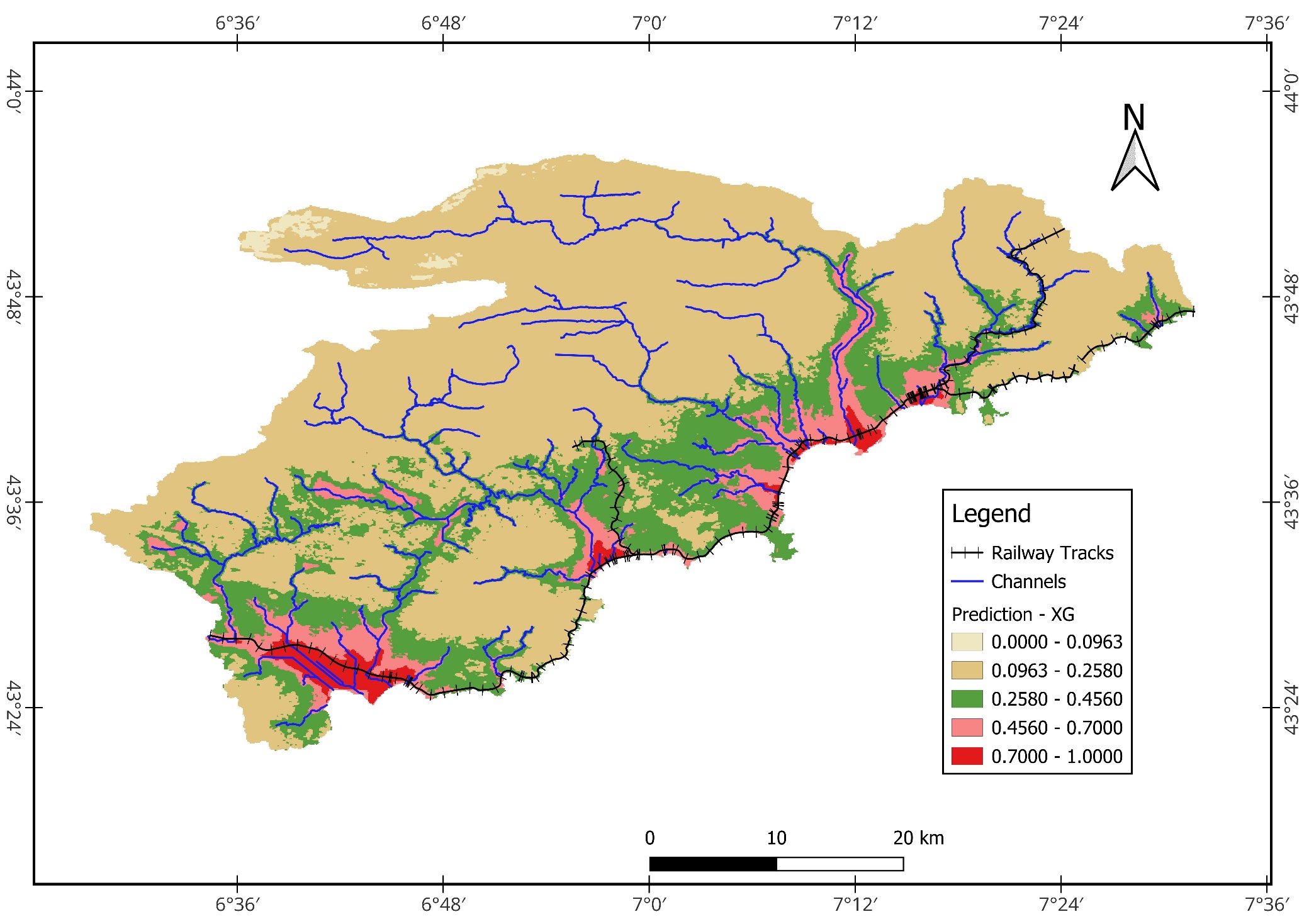}
  \caption{Flood susceptibility plot using the XGBoost model}
  \label{fig:xgb-susceptibility}
\end{figure}

\begin{figure}[htbp]
  \centering
  \includegraphics[width=\linewidth]{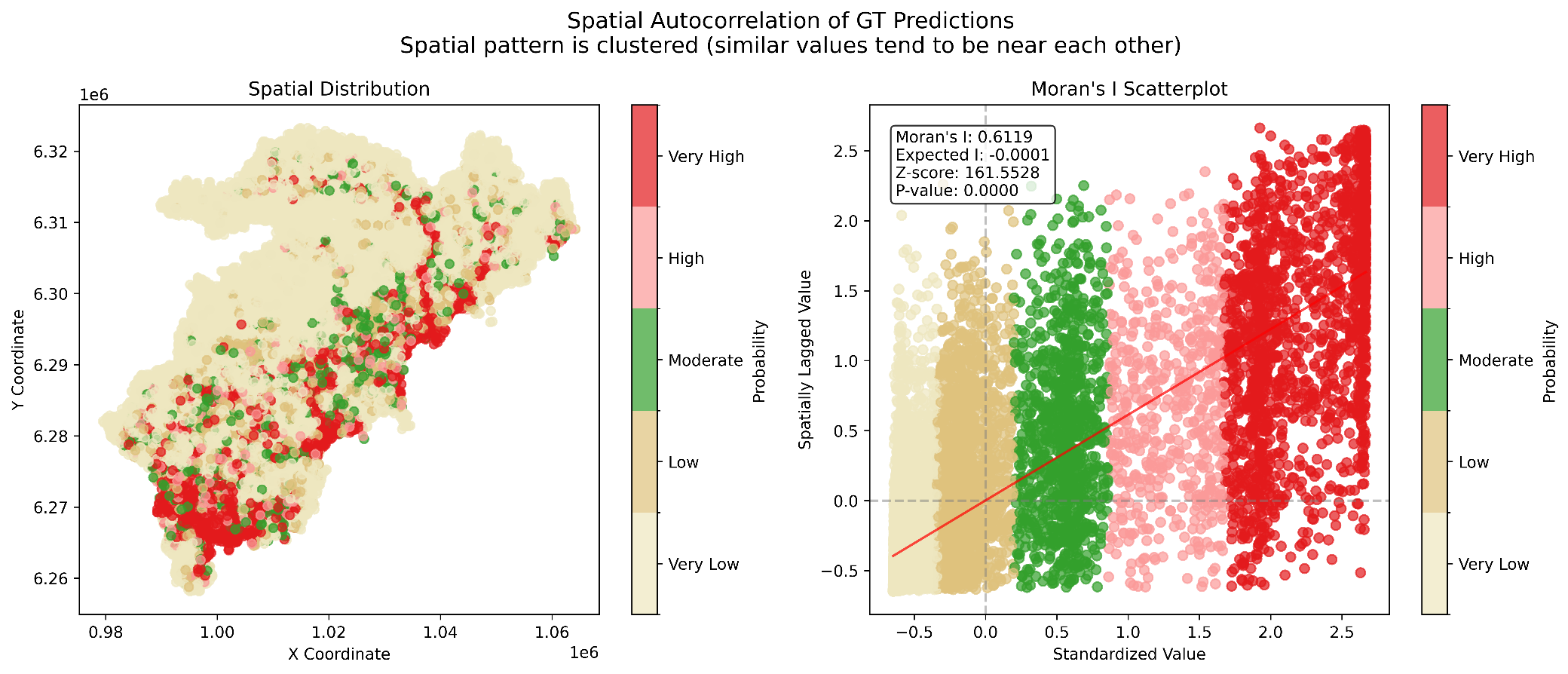}
  \caption{Moran's I spatial autocorrelation plot for the GT model}
  \label{fig:moran-gt}
\end{figure}

\begin{figure}[htbp]
  \centering
  \includegraphics[width=\linewidth]{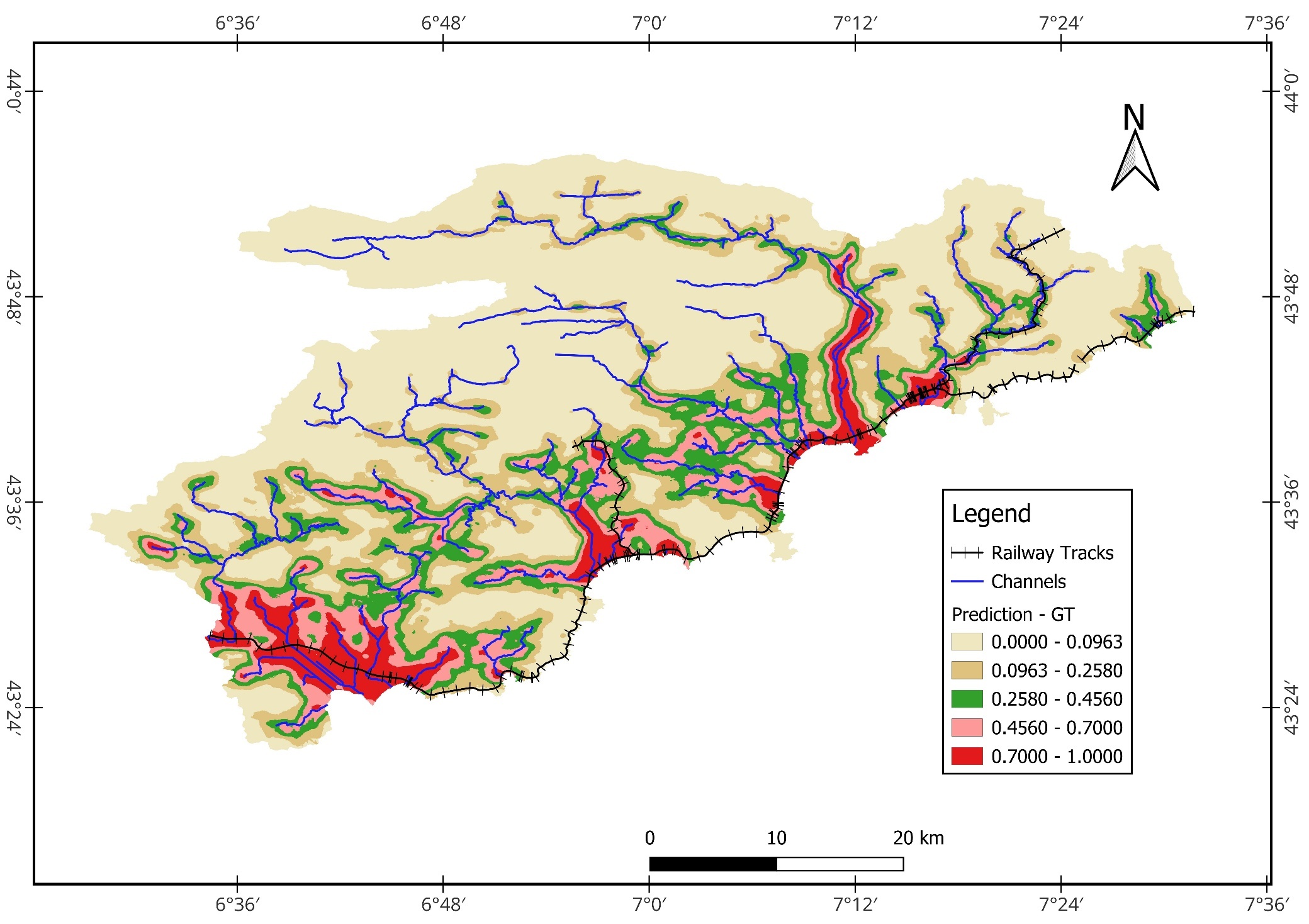}
  \caption{Flood susceptibility plot using the GT model}
  \label{fig:gt-susceptibility}
\end{figure}

\begin{figure}[htbp]
  \centering
  \includegraphics[width=\linewidth]{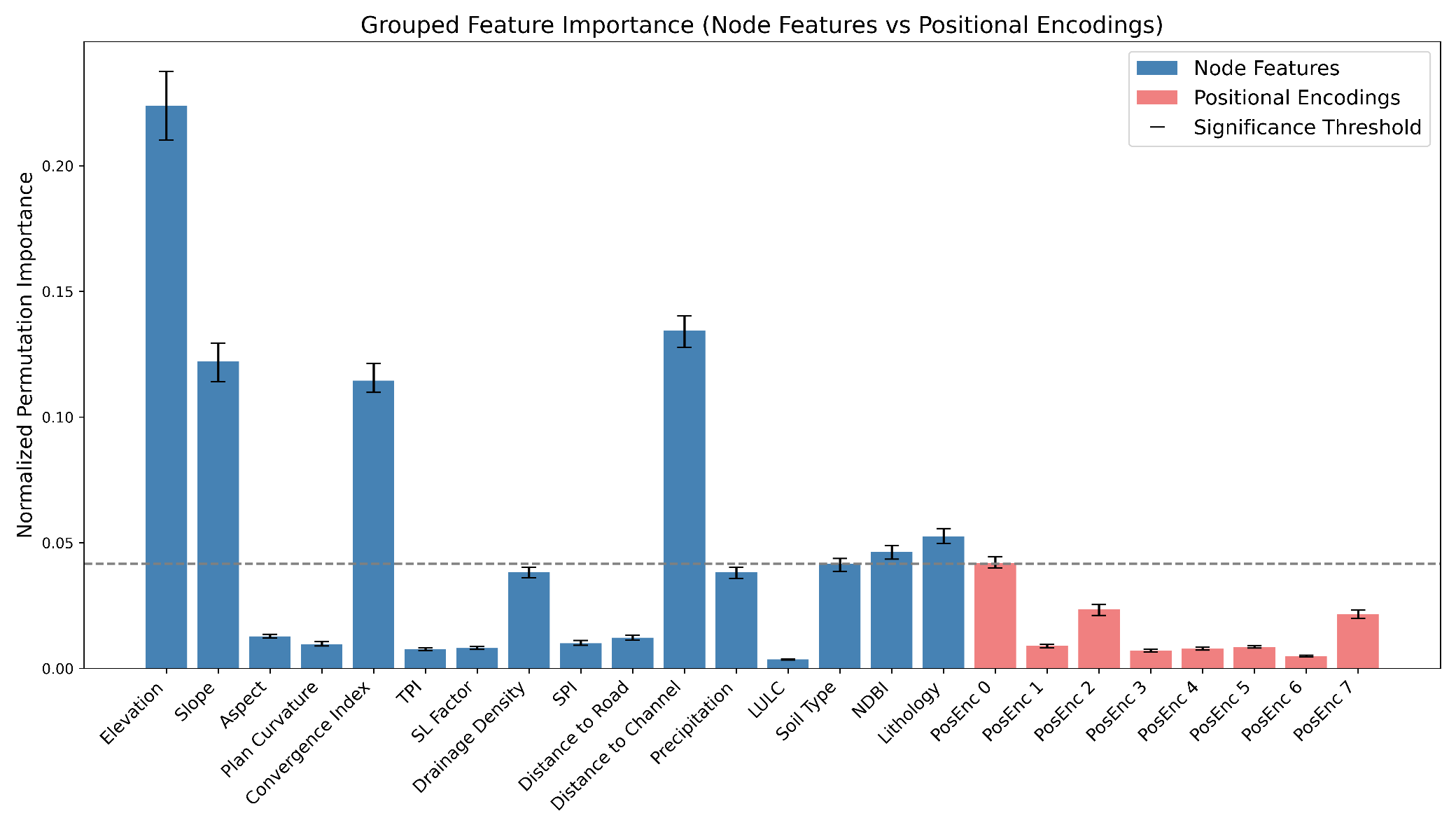}
  \caption{Global feature importance using the permutation method}
  \label{fig:perm-importance}
\end{figure}

\begin{figure}[htbp]
\centering
\captionsetup[sub]{justification=centering,singlelinecheck=false}
\begin{subfigure}[t]{0.49\textwidth}
  \centering
  \includegraphics[width=\linewidth]{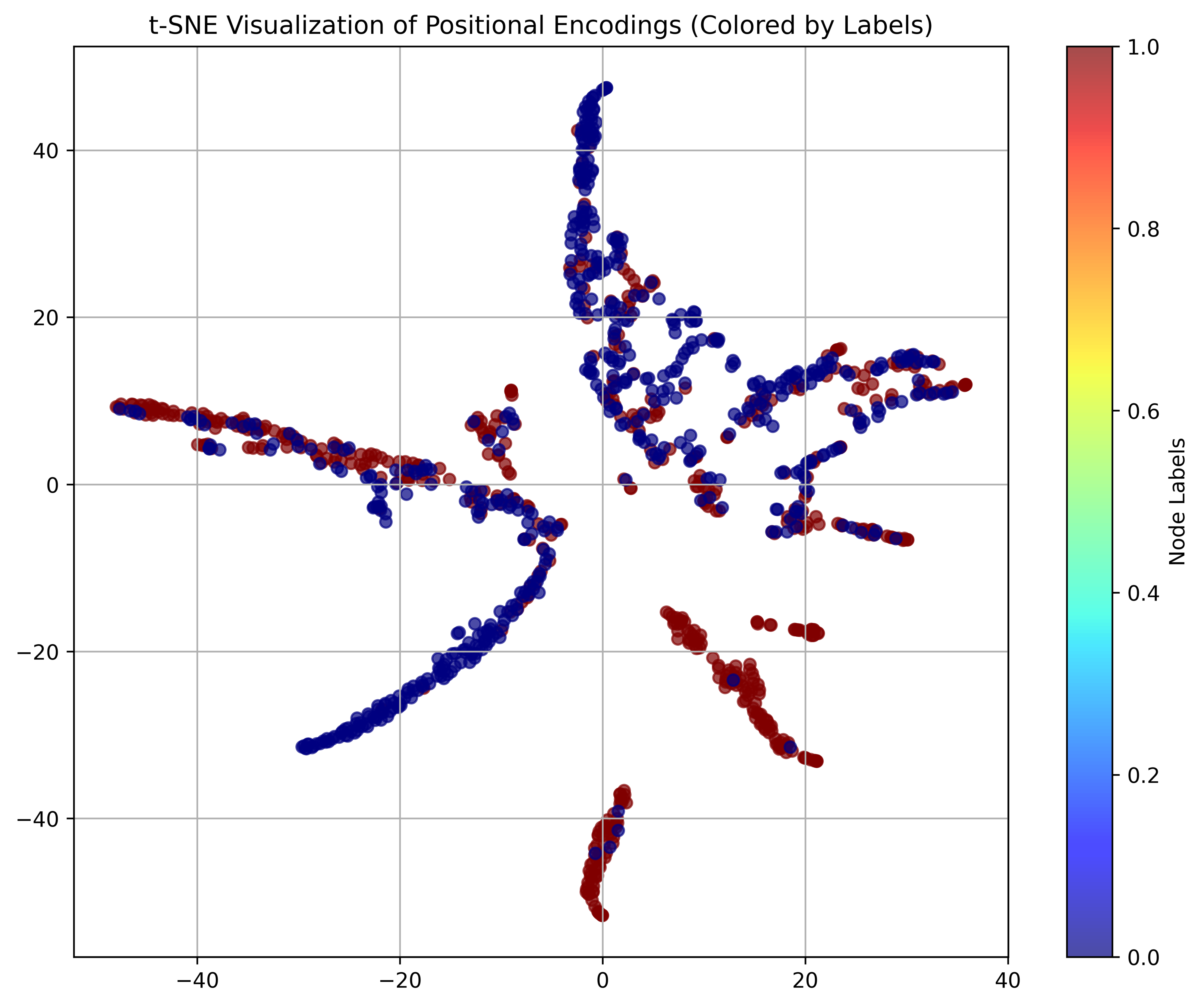}
  \caption{t-SNE}
  \label{fig:fig8-tsne}
\end{subfigure}\hfill
\begin{subfigure}[t]{0.49\textwidth}
  \centering
  \includegraphics[width=\linewidth]{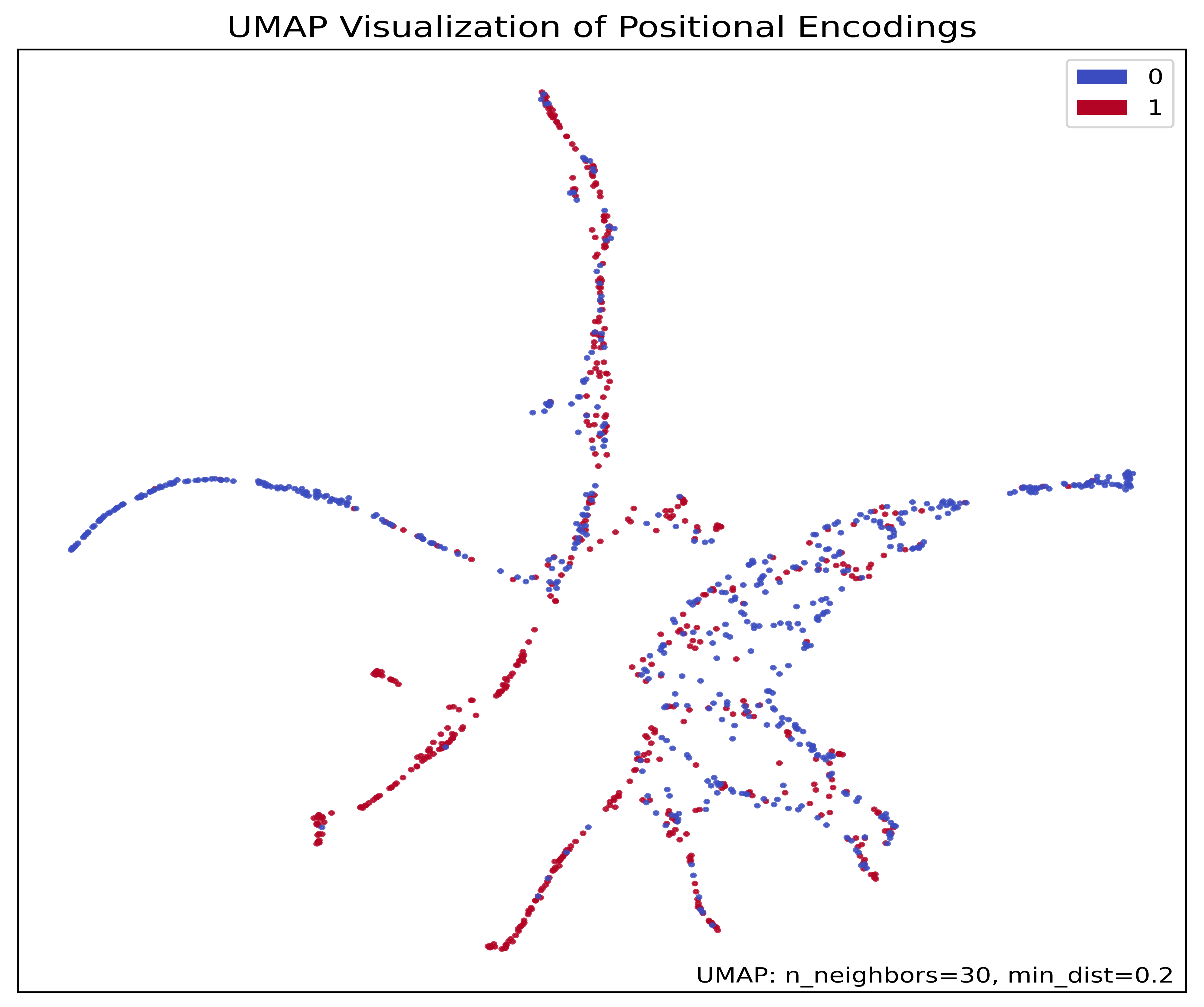}
  \caption{UMAP}
  \label{fig:fig8-umap}
\end{subfigure}
\caption{t-SNE (left) and UMAP (right) visualizations of the GT model positional encodings.}
\label{fig:fig8}
\end{figure}

\begin{figure}[htbp]
  \centering
  \includegraphics[width=\linewidth]{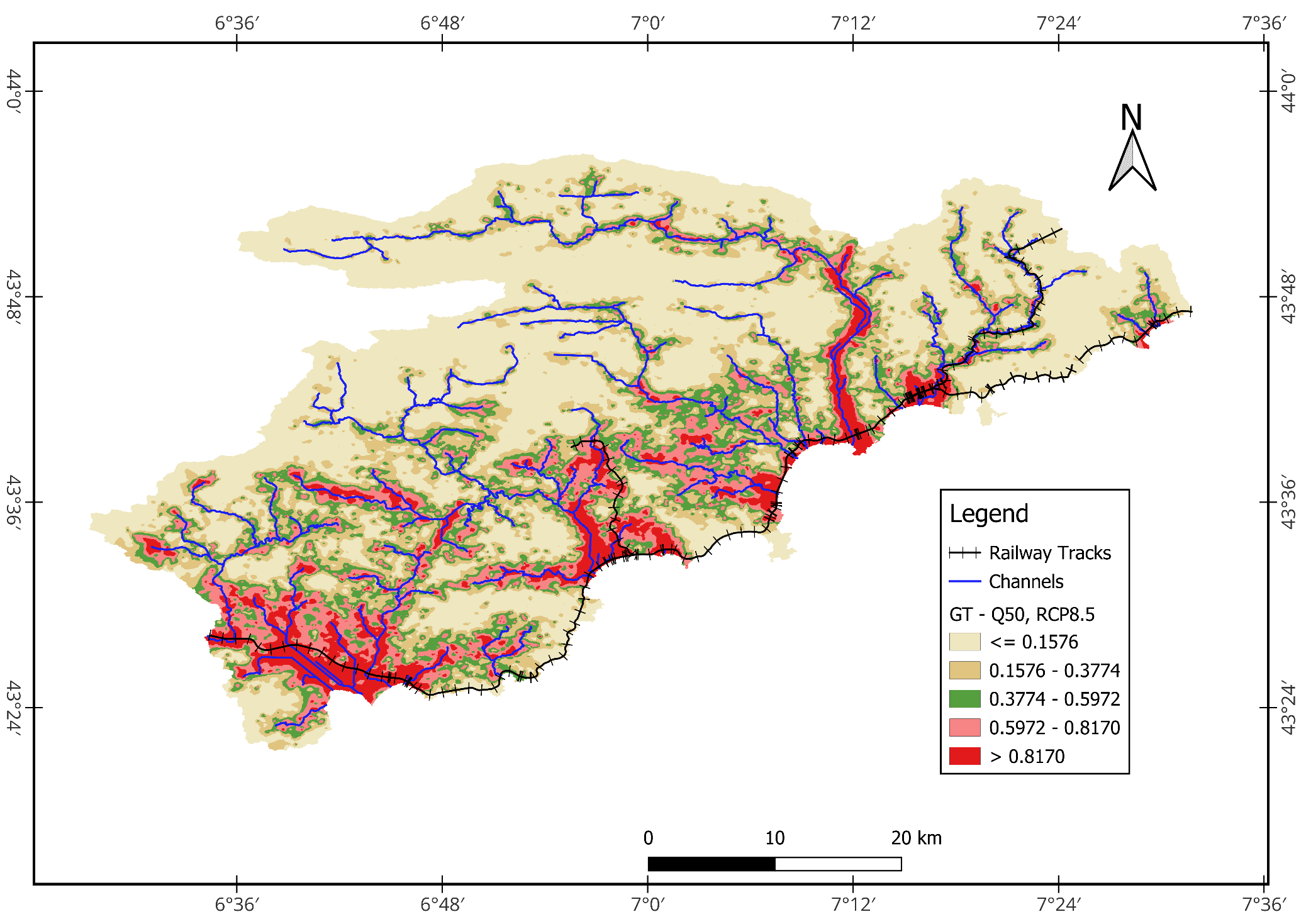}
  \caption{Projected flood susceptibility plot using the GT model for the year 2050, Q50 percentile under RCP 8.5 scenario}
  \label{fig:proj-q50}
\end{figure}

\begin{figure}[htbp]
  \centering
  \includegraphics[width=\linewidth]{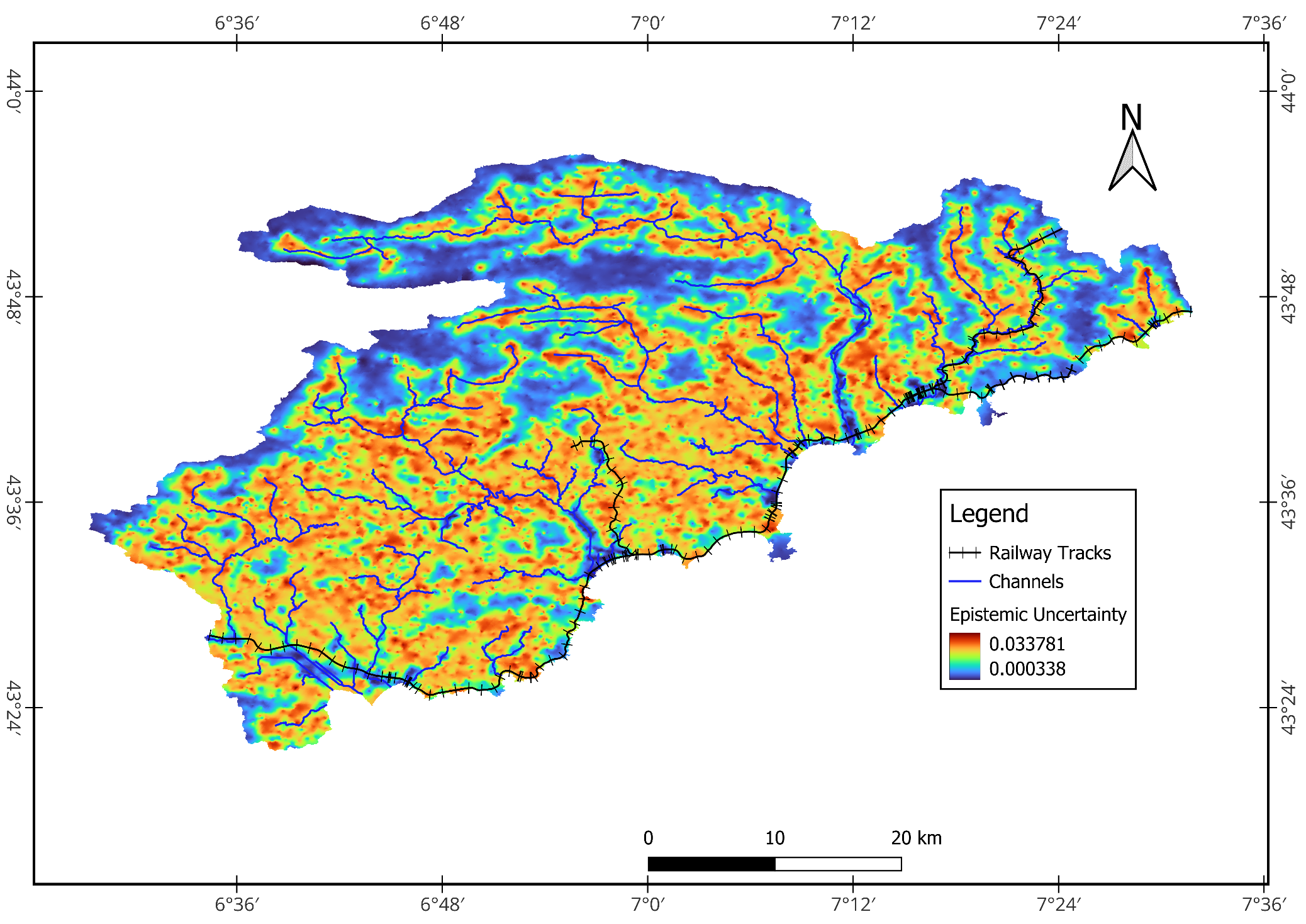}
  \caption{Epistemic uncertainty of the projected flood susceptibility model using the GT model for the year 2050, Q50 percentile under RCP 8.5 scenario}
  \label{fig:epis-uncer}
\end{figure}

\clearpage

\section{References}

Abedi, R., Costache, R., Shafizadeh-Moghadam, H., \& Pham, Q. B. (2022). Flash-flood susceptibility mapping based on XGBoost, random forest and boosted regression trees. Geocarto International.

Akiba, T., Sano, S., Yanase, T., Ohta, T., \& Koyama, M. (2019, July). Optuna: A next-generation hyperparameter optimization framework. In Proceedings of the 25th ACM SIGKDD international conference on knowledge discovery \& data mining (pp. 2623-2631). DOI: 10.48550/arXiv.1907.10902.

Arabameri, A., Seyed Danesh, A., Santosh, M., Cerda, A., Chandra Pal, S., Ghorbanzadeh, O., ... \& Chowdhuri, I. (2022). Flood susceptibility mapping using meta-heuristic algorithms. Geomatics, Natural Hazards and Risk, 13(1), 949-974. DOI: 10.1080/19475705.2022.2060138.

Belvederesi, G., Tanyas, H., Lipani, A., Dahal, A., \& Lombardo, L. (2025). Distribution-agnostic landslide hazard modelling via Graph Transformers. Environmental Modelling \& Software, 183, 106231. DOI: 10.1016/j.envsoft.2024.106231.

Bentivoglio, R., Isufi, E., Jonkman, S. N., \& Taormina, R. (2022). Deep learning methods for flood mapping: a review of existing applications and future research directions. Hydrology and Earth System Sciences Discussions, 2022, 1-50. DOI: 10.5194/hess-26-4345-2022, 2022.

Breiman, L. (2001). Random forests. Machine learning, 45, 5-32. DOI: 10.1023/A:1010933404324.

Bubeck, P., Dillenardt, L., Alfieri, L., Feyen, L., Thieken, A. H., \& Kellermann, P. (2019). Global warming to increase flood risk on European railways. Climatic Change, 155, 19-36. DOI: 10.1007/s10584-019-02434-5.

Bui, D. T., Tsangaratos, P., Ngo, P. T. T., Pham, T. D., \& Pham, B. T. (2019). Flash flood susceptibility modeling using an optimized fuzzy rule based feature selection technique and tree based ensemble methods. Science of the total environment, 668, 1038-1054. DOI: 10.1016/j.scitotenv.2019.02.422.

Bui, Q. T., Nguyen, Q. H., Nguyen, X. L., Pham, V. D., Nguyen, H. D., \& Pham, V. M. (2020). Verification of novel integrations of swarm intelligence algorithms into deep learning neural network for flood susceptibility mapping. Journal of Hydrology, 581, 124379. DOI: 10.1016/j.jhydrol.2019.124379.

Cheetham, M., Chirouze, F., \& Bredier, L. (2016). RISK VIP: Evaluation of Flood Risk on the French Railway Network Using an Innovative GIS Approach. In E3S Web of Conferences (Vol. 7, p. 10004). EDP Sciences. DOI: 10.1051/e3sconf/20160710004.

Chen, T., \& Guestrin, C. (2016, August). Xgboost: A scalable tree boosting system. In Proceedings of the 22nd ACM SIGKDD International Conference on Knowledge Discovery and Data Mining (pp. 785-794). DOI: 10.1145/2939672.2939785.

Chen, W., Li, Y., Xue, W., Shahabi, H., Li, S., Hong, H., ... \& Ahmad, B. B. (2020). Modeling flood susceptibility using data-driven approaches of naïve Bayes tree, alternating decision tree, and random forest methods. Science of The Total Environment, 701, 134979. DOI: 10.1016/j.scitotenv.2019.134979.

Choubin, B., Moradi, E., Golshan, M., Adamowski, J., Sajedi-Hosseini, F., \& Mosavi, A. (2019). An ensemble prediction of flood susceptibility using multivariate discriminant analysis, classification and regression trees, and support vector machines. Science of the Total Environment, 651, 2087-2096. DOI: 10.1016/j.scitotenv.2018.10.064.

Clark University. "Projected land cover in 2050" {[}Web Map{]}. Scale 1:50,000 and smaller. "ArcGIS Living Atlas of the World". April 2021. https://www.arcgis.com/home/item.html?id=3cce97cba8394287bcaf60f7618a5500 - (20 February 2025)

Cortes, C., \& Vapnik, V. (1995). Support-vector networks. Machine learning, 20, 273-297. DOI: 10.1007/BF00994018.

Conda-Forge Community. (2024). XGBoost: Scalable and Portable Gradient Boosting Library (Version 1.7.3) {[}Software{]}. Retrieved September 30, 2024, from https://anaconda.org/conda-forge/xgboost

Dwivedi, V. P., \& Bresson, X. (2020). A generalization of transformer networks to graphs. arXiv preprint arXiv:2012.09699.

Fenglin, W., Ahmad, I., Zelenakova, M., Fenta, A., Dar, M. A., Teka, A. H., ... \& Shafi, S. N. (2023). Exploratory regression modeling for flood susceptibility mapping in the GIS environment. Scientific Reports, 13(1), 247. DOI: 10.1038/s41598-023-27447-0.

Fisher, A., Rudin, C., \& Dominici, F. (2019). All models are wrong, but many are useful: Learning a variable\textquotesingle s importance by studying an entire class of prediction models simultaneously. Journal of Machine Learning Research, 20(177), 1-81. DOI: 10.48550/arXiv.1801.01489.

Gao, W., Liao, Y., Chen, Y., Lai, C., He, S., \& Wang, Z. (2024). Enhancing transparency in data-driven urban pluvial flood prediction using an explainable CNN model. Journal of Hydrology, 645, 132228. DOI: 10.1016/j.jhydrol.2024.132228.

Ghosh, S., Saha, S., \& Bera, B. (2022). Flood susceptibility zonation using advanced ensemble machine learning models within Himalayan foreland basin. Natural Hazards Research, 2(4), 363-374. DOI: 10.1016/j.nhres.2022.06.003.

Gilmer, J., Schoenholz, S. S., Riley, P. F., Vinyals, O., \& Dahl, G. E. (2017, July). Neural message passing for quantum chemistry. In International conference on machine learning (pp. 1263-1272). PMLR. DOI: 10.48550/arXiv.1704.01212.

Ginesta, M., Yiou, P., Messori, G., \& Faranda, D. (2023). A methodology for attributing severe extratropical cyclones to climate change based on reanalysis data: the case study of storm Alex 2020. Climate Dynamics, 61(1), 229-253. DOI: 10.1007/s00382-022-06565-x.

Grinsztajn, L., Oyallon, E., \& Varoquaux, G. (2022). Why do tree-based models still outperform deep learning on typical tabular data? Advances in neural information processing systems, 35, 507-520.

Habibi, A., Delavar, M. R., Nazari, B., Pirasteh, S., \& Sadeghian, M. S. (2023). A novel approach for flood hazard assessment using hybridized ensemble models and feature selection algorithms. International Journal of Applied Earth Observation and Geoinformation, 122, 103443. DOI: 10.1016/j.jag.2023.103443.

Hong, H., Tsangaratos, P., Ilia, I., Liu, J., Zhu, A. X., \& Chen, W. (2018). Application of fuzzy weight of evidence and data mining techniques in construction of flood susceptibility map of Poyang County, China. Science of the total environment, 625, 575-588. DOI: 10.1016/j.scitotenv.2017.12.256.

Islam, A. R. M. T., Talukdar, S., Mahato, S., Kundu, S., Eibek, K. U., Pham, Q. B., ... \& Linh, N. T. T. (2021). Flood susceptibility modelling using advanced ensemble machine learning models. Geoscience Frontiers, 12(3), DOI: 101075.10.1016/j.gsf.2020.09.006.

Janizadeh, S., Pal, S. C., Saha, A., Chowdhuri, I., Ahmadi, K., Mirzaei, S., ... \& Tiefenbacher, J. P. (2021). Mapping the spatial and temporal variability of flood hazard affected by climate and land-use changes in the future. Journal of Environmental Management, 298, 113551. DOI: 10.1016/j.jenvman.2021.113551.

Karra, K., Kontgis, C., Statman-Weil, Z., Mazzariello, J. C., Mathis, M., \& Brumby, S. P. (2021, July). Global land use/land cover with Sentinel 2 and deep learning. In 2021 IEEE International Geoscience and Remote Sensing Symposium IGARSS (pp. 4704-4707). IEEE. DOI:10.1109/IGARSS47720.2021.9553499.

Khosravi, K., Shahabi, H., Pham, B. T., Adamowski, J., Shirzadi, A., Pradhan, B., ... \& Prakash, I. (2019). A comparative assessment of flood susceptibility modeling using multi-criteria decision-making analysis and machine learning methods. Journal of Hydrology, 573, 311-323. DOI: 10.1016/j.jhydrol.2019.03.073.

Koks, E. E., Rozenberg, J., Zorn, C., Tariverdi, M., Vousdoukas, M., Fraser, S. A., ... \& Hallegatte, S. (2019). A global multi-hazard risk analysis of road and railway infrastructure assets. Nature communications, 10(1), 2677. DOI: 10.1038/s41467-019-10442-3.

Lundberg, S. M., \& Lee, S. I. (2017). A unified approach to interpreting model predictions. Advances in neural information processing systems, 30. DOI: 10.48550/arXiv.1705.07874.

Geary, R. C. (1954). The contiguity ratio and statistical mapping. The incorporated statistician, 5(3), 115-146. https://doi.org/10.2307/2986645.

Gharakhanlou, N., \& Perez, L. (2022). Spatial prediction of current and future flood susceptibility: examining the implications of changing climates on flood susceptibility using machine learning models. Entropy, 24(11), 1630. DOI: 10.3390/e24111630.

Meinshausen, M., Smith, S. J., Calvin, K., Daniel, J. S., Kainuma, M. L., Lamarque, J. F., ... \& van Vuuren, D. P. (2011). The RCP greenhouse gas concentrations and their extensions from 1765 to 2300. Climatic change, 109, 213-241. DOI: 10.1007/s10584-011-0156-z.

Météo-France, IPSL, CERFACS, CNRM-GAME. (2025). DRIAS: French regionalized climate projections Dataset{]}. Retrieved from https://www.drias-climat.fr/ on February 22, 2025.

Montoya-Araque, E., Montoya-Noguera, S., Lopez‐caballero, F., \& Gatti, F. (2025). Numerical earthquake-induced landslide hazard assessment at regional scale in the Colombian Andes. Soil Dynamics and Earthquake Engineering. DOI: 10.1016/j.soildyn.2025.109370

Moran, P. A. (1950). Notes on continuous stochastic phenomena. Biometrika, 37(1/2), 17-23. DOI: 10.2307/2332142.

Mosavi, A., Ozturk, P., \& Chau, K. W. (2018). Flood prediction using machine learning models: Literature review. Water, 10(11), 1536. DOI: 10.3390/w10111536.

Nguyen, H. D. (2023). Spatial modeling of flood hazard using machine learning and GIS in Ha Tinh province, Vietnam. Journal of Water and Climate Change, 14(1), 200-222. DOI: 10.2166/wcc.2022.257.

Nguyen, H. D., Nguyen, Q. H., Dang, D. K., Van, C. P., Truong, Q. H., Pham, S. D., ... \& Petrisor, A. I. (2024). A novel flood risk management approach based on future climate and land use change scenarios. Science of The Total Environment, 921, 171204. DOI: 10.1016/j.scitotenv.2024.171204.

Paszke, A., Gross, S., Massa, F., Lerer, A., Bradbury, J., Chanan, G., ... \& Chintala, S. (2019). Pytorch: An imperative style, high-performance deep learning library. Advances in neural information processing systems, 32. DOI: 10.48550/arXiv.1912.01703.

Pedregosa, F., Varoquaux, G., Gramfort, A., Michel, V., Thirion, B., Grisel, O., ... \& Duchesnay, É. (2011). Scikit-learn: Machine learning in Python. The Journal of machine Learning research, 12, 2825-2830.

Pham, B. T., Phong, T. V., Nguyen, H. D., Qi, C., Al-Ansari, N., Amini, A., ... \& Tien Bui, D. (2020). A comparative study of kernel logistic regression, radial basis function classifier, multinomial naïve bayes, and logistic model tree for flash flood susceptibility mapping. Water, 12(1), 239. DOI: 10.3390/w12010239.

Pradhan, B., Lee, S., Dikshit, A., \& Kim, H. (2023). Spatial flood susceptibility mapping using an explainable artificial intelligence (XAI) model. Geoscience Frontiers, 14(6), 101625. DOI: 10.1016/j.gsf.2023.101625.

Prakash, N., \& Manconi, A. (2021, July). Rapid Mapping of Landslides Triggered by the Storm Alex, October 2020. In 2021 IEEE International Geoscience and Remote Sensing Symposium IGARSS (pp. 1808-1811). IEEE. DOI: 10.1109/IGARSS47720.2021.9553321.

Rogers, J. S., Maneta, M. M., Sain, S. R., Madaus, L. E., \& Hacker, J. P. (2025). The role of climate and population change in global flood exposure and vulnerability. Nature Communications, 16(1), 1287. DOI: 10.1038/s41467-025-56654-8.

Rudin, C. (2019). Stop explaining black box machine learning models for high stakes decisions and use interpretable models instead. Nature Machine Intelligence, 1(5), 206-215.

Seabold, S., \& Perktold, J. (2010). Statsmodels: econometric and statistical modeling with Python. SciPy, 7(1), 92-96.

Toribio Diaz, C., \& Vallhonrat Blanco, P. J. (2025). Resilient railways facing heavy rains (Report No. 5-25006E). UIC. https://shop.uic.org/en/other-reports/14853-resilient-railways-facing-heavy-rains.html

Trenberth, K. E. (2011). Changes in precipitation with climate change. Climate Research, 47(1-2), 123-138. DOI: 10.3354/cr00953.

Vaswani, A., Shazeer, N., Parmar, N., Uszkoreit, J., Jones, L., Gomez, A. N., ... \& Polosukhin, I. (2017). Attention is all you need. Advances in neural information processing systems, 30.

Van der Maaten, L., \& Hinton, G. (2008). Visualizing data using t-SNE. Journal of machine learning research, 9(11).

Veličković, P., Cucurull, G., Casanova, A., Romero, A., Lio, P., \& Bengio, Y. (2017). Graph attention networks. arXiv preprint arXiv:1710.10903.

Wang, Z., Lyu, H., \& Zhang, C. (2023). Pluvial flood susceptibility mapping for data-scarce urban areas using graph attention network and basic flood conditioning factors. Geocarto International, 38(1), 2275692. DOI: 10.1080/10106049.2023.2275692.

Wang, Z., Lyu, H., \& Zhang, C. (2024). Urban pluvial flood susceptibility mapping based on a novel explainable machine learning model with synchronous enhancement of fitting capability and explainability. Journal of Hydrology, 642, 131903. DOI: 10.1016/j.jhydrol.2024.131903.

Widya, L. K., Rezaie, F., Lee, W., Lee, C. W., Nurwatik, N., \& Lee, S. (2024). Flood susceptibility mapping of Cheongju, South Korea based on the integration of environmental factors using various machine learning approaches. Journal of Environmental Management, 364, 121291.

Zhang, Q., He, Y., Zhang, Y., Lu, J., Zhang, L., Huo, T., ... \& Zhang, Y. (2024). A Graph-Transformer method for landslide susceptibility mapping. IEEE Journal of Selected Topics in Applied Earth Observations and Remote Sensing. DOI: 10.1109/JSTARS.2024.3437751.

Zhao, G., Pang, B., Xu, Z., Cui, L., Wang, J., Zuo, D., \& Peng, D. (2021). Improving urban flood susceptibility mapping using transfer learning. Journal of Hydrology, 602, 126777. DOI: 10.1016/j.jhydrol.2021.126777.

Zhao, G., Pang, B., Xu, Z., Peng, D., \& Xu, L. (2019). Assessment of urban flood susceptibility using semi-supervised machine learning model. Science of the Total Environment, 659, 940-949. DOI: 10.1016/j.scitotenv.2018.12.217.

Zhou, J., Cui, G., Hu, S., Zhang, Z., Yang, C., Liu, Z., ... \& Sun, M. (2020). Graph neural networks: A review of methods and applications. AI open, 1, 57-81.
\clearpage

\includepdf[pages=-]{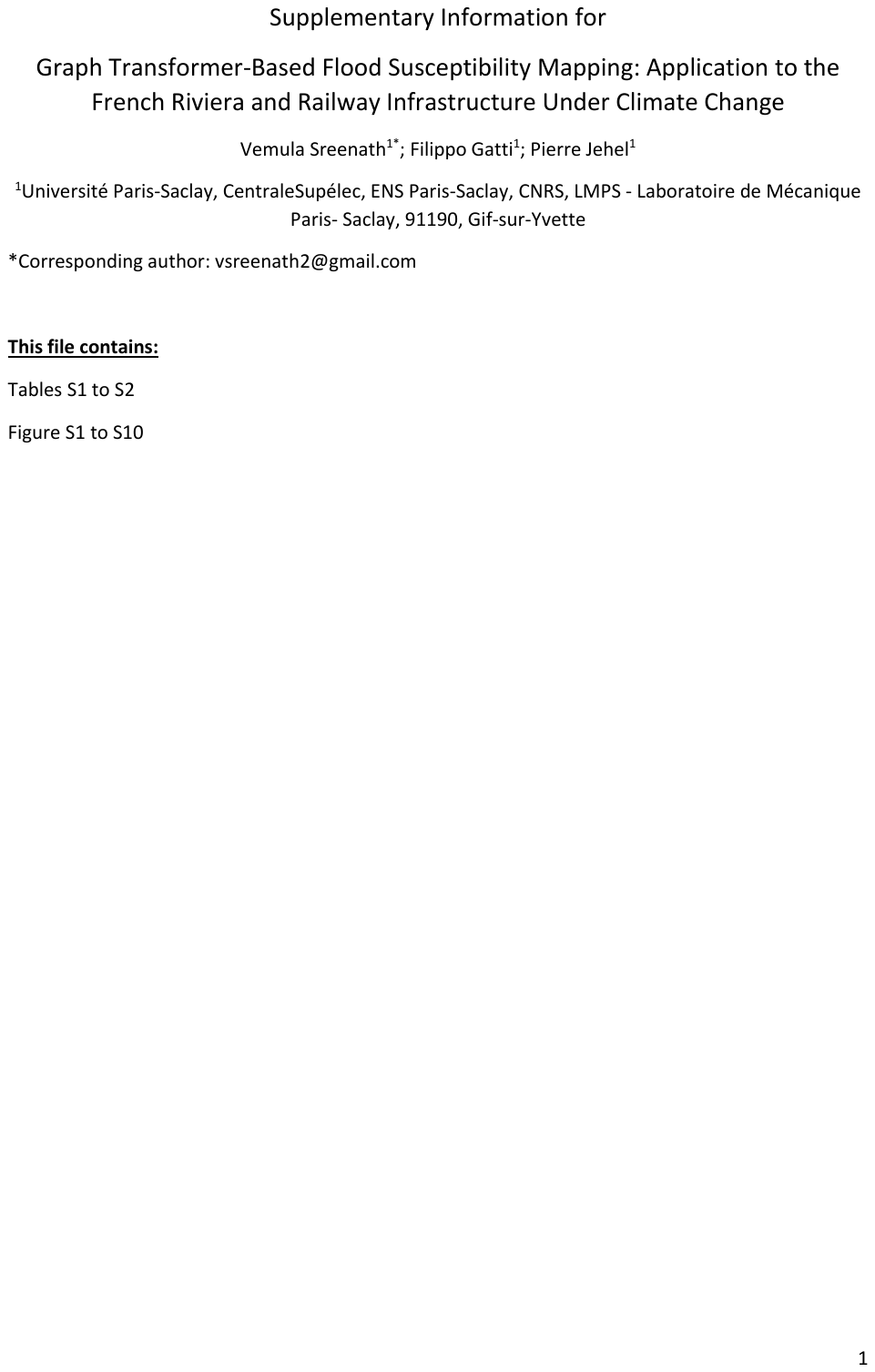}

\end{document}